\documentclass[twocolumn,mathlines]{aastex631}
\usepackage{xcolor}
\usepackage{apjfonts}
\usepackage{graphicx}
\usepackage{epstopdf}
\usepackage{ulem}
\usepackage[utf8]{inputenc}
\usepackage{booktabs,tabularx}
\usepackage[encapsulated]{CJK}
\usepackage[T1]{fontenc}
\usepackage{booktabs}
\usepackage{threeparttable}
\usepackage{mdwlist}
\usepackage{amsmath} 
\usepackage{enumitem}

\begin{document}
\newcommand{\ls}{{_<\atop^{\sim}}}
\newcommand{\gs}{{_>\atop^{\sim}}}
\def \spose#1{\hbox  to 0pt{#1\hss}}  
\def \ls{\mathrel{\spose{\lower 3pt\hbox{$\sim$}}\raise  2.0pt\hbox{$<$}}}
\def \gs{\mathrel{\spose{\lower  3pt\hbox{$\sim$}}\raise 2.0pt\hbox{$>$}}}
\newcommand{\Ha}{\hbox{{\rm H}$\alpha$}}
\newcommand{\Hb}{\hbox{{\rm H}$\beta$}}
\newcommand{\Ovi}{\hbox{{\rm O}\kern 0.1em{\sc vi}}}
\newcommand{\OIII}{\hbox{[{\rm O}\kern 0.1em{\sc iii}]}}
\newcommand{\OII}{\hbox{[{\rm O}\kern 0.1em{\sc ii}]}}
\newcommand{\lya}{Ly$\alpha$}
\newcommand{\NII}{\hbox{[{\rm N}\kern 0.1em{\sc ii}]}}
\newcommand{\SII}{\hbox{[{\rm S}\kern 0.1em{\sc ii}]}}
\newcommand{\angstrom}{\textup{\AA}}
\newcommand\ionn[2]{#1$\;${\scshape{#2}}}
\newcommand{\fesc}{$ f_{\rm esc}$}
\newcommand{\flya}{$ f_{\rm esc}^{Ly\alpha}$}



\title{Lyman Continuum leakage from massive leaky starbursts: A different class of emitters?}

\shorttitle{Lyman Continuum escape }
 
\shortauthors{Roy et al.}


\correspondingauthor{Namrata Roy}
\email{namratar@asu.edu}

\author[0000-0002-4430-8846]{Namrata Roy}
\affiliation{William H. Miller III Department of Physics and Astronomy, Johns Hopkins University, Baltimore, MD, 21218}
\affiliation{School of Earth and Space Exploration, Arizona State University, Tempe, AZ, 85287}

\author[0000-0001-6670-6370]{Timothy Heckman}
\affiliation{William H. Miller III Department of Physics and Astronomy, Johns Hopkins University, Baltimore, MD, 21218}
\affiliation{School of Earth and Space Exploration, Arizona State University, Tempe, AZ, 85287}

\author[0000-0002-6586-4446]{Alaina Henry}
\affiliation{Space Telescope Science Institute, 3700 San Martin Drive, Baltimore MD, 21218} 
\affiliation{Center for Astrophysical Sciences, Department of Physics and Astronomy, Johns Hopkins University, Baltimore, MD, 21218}

\author[0000-0002-0302-2577]{John Chisholm}
\affiliation{Department of Astronomy, University of Texas at Austin, 2515 Speedway, Austin, Texas 78712, USA}

\author[0000-0002-0159-2613]{Sophia Flury}
\affiliation{Institute for Astronomy, University of Edinburgh Royal Observatory, Blackford Hill, Edinburgh, EH9 3HJ, UK}

\author[0000-0003-2685-4488]{Claus Leitherer}
\affiliation{Space Telescope Science Institute, 3700 San Martin Drive, Baltimore MD, 21218} 

\author[0000-0001-8587-218X]{Matthew J. Hayes}
\affiliation{Department of Astronomy, Oskar Klein Centre, Stockholm University, AlbaNova University Center, 10691 Stockholm, Sweden}

\author[0000-0002-6790-5125]{Anne Jaskot}
\affiliation{Department of Astronomy, Williams College, Williamstown, MA 01267, USA}

\author[0000-0001-7673-2257]{Zhiyuan Ji}
\affiliation{Steward Observatory, University of Arizona, 933 N. Cherry Avenue, Tucson, AZ 85721, USA}

\author[0000-0001-7144-7182]{Daniel Schaerer}
\affiliation{Observatoire de Gen\`eve, Universit\'e de Gen\`eve, Chemin Pegasi 51, 1290 Versoix, Switzerland} 
\affiliation{CNRS, IRAP, 14 Avenue E. Belin, 31400 Toulouse, France}

\author[0000-0001-9269-5046]{Bingjie Wang (\begin{CJK*}{UTF8}{gbsn}王冰洁\ignorespacesafterend\end{CJK*})}
\affiliation{Department of Astronomy \& Astrophysics, The Pennsylvania State University, University Park, PA 16802, USA}
\affiliation{Institute for Computational \& Data Sciences, The Pennsylvania State University, University Park, PA 16802, USA}
\affiliation{Institute for Gravitation and the Cosmos, The Pennsylvania State University, University Park, PA 16802, USA}

\author[0000-0002-2724-8298]{Sanchayeeta Borthakur}
\affiliation{School of Earth and Space Exploration, Arizona State University, 781 Terrace Mall, Tempe, AZ 85287, USA}

\author[0000-0002-9217-7051]{Xinfeng Xu}
\affiliation{Department of Physics and Astronomy, Northwestern University,
2145 Sheridan Road, Evanston, IL, 60208, USA.}
\affiliation{Center for Interdisciplinary Exploration and Research in
Astrophysics (CIERA), 1800 Sherman Avenue,
Evanston, IL, 60201, USA.}

\author[0000-0002-3005-1349]{Göran Östlin}
\affiliation{Department of Astronomy, Oskar Klein Centre, Stockholm University, AlbaNova University Center, 10691 Stockholm, Sweden}


\begin{abstract}

The origin of Lyman continuum (LyC) photons responsible for reionizing the universe remains largely unknown, with the fraction of escaping LyC photons from galaxies at $z \sim 6$ to 12 still uncertain. Direct detection of LyC photons from this epoch is challenging due to intergalactic medium absorption, making lower-redshift analogs valuable for studying LyC leakage.
In this study, we present Hubble Space Telescope Cosmic Origins Spectrograph (HST COS) observations of five low-redshift (z $\sim$ 0.3) massive starburst galaxies, selected for high stellar mass and weak [S II] nebular emission, an indirect tracer of LyC escape. LyC leakage is detected in three of the five galaxies, highlighting weak [S II] as a reliable tracer — a finding supported by recent JWST discoveries of z $>$ 5 galaxies with similarly weak [S II] emission. The dust-corrected LyC escape fractions ($\rm f_{esc, HI}$), representing LyC photons that would escape without dust, range from 33\% to 84\%. However, the absolute escape fractions ($\rm f_{esc, tot}$), accounting for both neutral hydrogen absorption and dust attenuation, are substantially lower, between 1\% and 3\%. This indicates that, although these galaxies are nearly optically thin to HI, their significant dust content restricts LyC escape. These [S II]-weak, massive leakers differ from typical low-redshift LyC emitters, exhibiting higher metallicity, lower ionization states, greater dust extinction, and higher star formation surface densities. We suggest that feedback-driven winds in these compact starbursts generate ionized channels, allowing LyC escape in line with a "picket-fence" model, indicating a distinct mechanism for LyC leakage.

\end{abstract}



\section{Introduction} \label{sec:Introduction}


The epoch of Reionization (EoR) is a critical phase in the history of the Universe when the intergalactic medium (IGM) changes its state from neutral to completely ionized \citep[see][ for a review]{robertson22}. It also signifies the period when the first generation of massive stars and black holes form. Observations from cosmic microwave background (CMB), WMAP and Ly$\alpha$ forest absorption troughs in distant quasars bear evidence that the major part of the reionization occurred as early as z$\sim$ 12, and lasted until z $\sim$ 6 \citep[e.g., ][]{becker01, fan06, banados18, mason18, planck20}. But what are the primary sources for reionization?  The early star forming galaxies are hypothesized to be the best candidates to supply the ionizing photons necessary for the Universe to be ionized \citep{bouwens16}.  However, the fraction of these ionizing Lyman continuum (LyC) photons that are actually able to escape from dense neutral gas into the IGM is poorly constrained. An average LyC escape fraction (\fesc) of 20\% or higher is required for star forming galaxies to be the major source for reionization \citep{bouwens11, robertson15}, although recent results show that the required \fesc \ do not need to be that high \citep{finkelstein19, atek24}.  It is however crucial to find star forming galaxies with high LyC escape to confirm their role in reionization.


Although many attempts have been made to observationally constrain the escape fraction in star forming galaxies across a wide range of redshift, most of the early observations yielded upper limits only with a few detections, indicating \fesc\ $<$ 3\% \citep{leitherer95, deharveng01, malkan03, alavi20}. With a more extensive search program in recent years using the Hubble Space Telescope (HST) Cosmic Origins Spectrograph \citep[COS; ][]{green12}, convincing detections of LyC leaking emission have been found in a few tens of sources, mostly from local (0.2$<$z$<$0.4) starbursts, ``Green Pea'' galaxies, and Lyman break analogs \citep{heckman05, hoopes07, iwata09, borthakur14, izotov16, izotov16b, izotov18, izotov18b, izotov21, wang19, wang21, flury22, xu22}. Notably, the recently completed Low Redshift Lyman Continuum Survey  \citep[LzLCS+; ][]{saldana-lopez22, flury22, flury22b} assembled a sample of 66 star forming LyC leaking ``candidate'' galaxies from the Sloan Digital sky survey \citep[SDSS:][]{york00}  and Galaxy Evolution Explorer \citep[GALEX:][]{martin05}, which were observed with HST/COS. The program added 35 new LyC detections with 97.7\% confidence. These Lyman continuum emitters (LCEs) serve as the largest Lyman Continuum Emitters sample detected to date. In addition, a few Lyman continuum detections with high escape fractions have also been reported at high redshifts \citep{vanzella18, fletcher19, vanzella19}.


As the confirmed direct detections of LyC escape have increased over the last couple of years, correlations have emerged between \fesc \ and their host galaxy properties. For example, there is a clear trend for leaky galaxies with high \fesc\ to preferentially have low stellar mass ($< \rm 10^{9} \ M_{\odot}$), high star formation rate (SFR), low metallicities ($<$ 1/3 solar) strong nebular emission lines, highly ionized gas  ([OIII]5007/[OII]3727 $>$ 3), low-dust attenuation and strong Ly$\alpha$ emission \citep{overzier08, overzier10, izotov16, izotov18, flury22b, saldana-lopez22, jaskot24, chisholm22}.  These signposts have been used as indirect tracers of escaping LyC emission and have been combined recently into a multivariate indicator of LyC escape \citep{jaskot24, jaskot24b}. 
%


It is imperative to consider the physical conditions that facilitate the escape of sufficient ionizing photons into the IGM, and result in the observed correlations discussed above \citep[e.g. as discussed in ][]{steidel18, gazagnes20, saldana-lopez22, flury22b}. Hot massive stars, the primary source of the ionizing radiation, reside in extremely gas rich regions in star forming galaxies with a very high Hydrogen column density  ($\rm 10^{21}$ to $\rm 10^{24} \ cm^{-2} $). This is roughly 4 to 7 orders of magnitude higher than the HI column density required to produce optical depth of unity at the Lyman edge ($\sim \rm 10^{17} \ cm^{-2}$). Thus extreme conditions in the interstellar medium (ISM) are necessary for a significant fraction of the ionizing flux to escape the IGM. 
One possible scenario is that ionizing radiation field from intense star forming regions can almost fully photoionize the surroundings, and thus can create channels of ionized gas that enables the escape of LyC photons. This ``density-bounded nebula'' picture is consistent with higher LyC escape found in galaxies with higher ionization, strong nebular emission lines, and low stellar masses with intense starburst activities.

In contrast, \cite{borthakur14} and \cite{wang19} found four galaxies which appear to be an entirely different class of LyC leaking sources. These galaxies exhibit properties entirely different from the traditionally known LCEs and are characterized by large stellar mass $\rm > 10^{10} \ M_{\odot} $, a much lower ionization state ([OIII]/[OII] $\sim$ 1), with the UV emission dominated by an extremely compact (few hundred parsecs) central region. They also display a relative weakness of [SII]6717\AA, 6731 \AA \  lines defined with respect to typical star forming galaxies. A deficit in [S II] emission has been proposed as an indicator of galaxies that are optically thin to ionizing radiation. Recent JWST observations have revealed several high-redshift ($z > 5$) galaxies with extremely weak [S II] lines \citep{sanders23, cameron23, hayes25}. In particular, \cite{hayes25} show that the most highly ionized high-$z$ galaxies, when stacked, frequently display a pronounced [S II] deficit and emerge as outliers in multiple properties, including $\beta$ slope, Mg II escape fraction, and ionizing photon production efficiency. These characteristics closely resemble those of the [S II]-deficient Lyman continuum leakers reported by \citet{wang19}.

It has been argued that a different mechanism may be at play here: these massive and compact starbursts drive extreme galactic winds, which blow holes and clear out low density channels in the neutral ISM through which LyC photons can escape \citep[e.g.,][]{bergvall06, heckman11}. In a classical HII region, a [SII] deficiency indicates a depleted partially ionized region near the outer edge of the Stromgren sphere \citep{pellegrini12, borthakur14, heckman11, alexandroff15}. In the massive, compact leaky galaxies this is possibly due to the fully ionized channels created by the feedback driven outflows.


Thus, there may be two different mechanisms to create channels in the ISM to enable LyC escape. A third possibility has also been proposed, in which merger-driven interactions disrupt the ISM and produce such low-density channels \citep{lereste24, yuan21, zhu24, mascia24}. However, we do not explore this scenario here, as detailed morphological analyses and merger identification for our sources lie beyond the scope of this work.
In this study, we expand the second sample of members of the massive, compact leaky starbursts by observing five new candidates using HST/COS observations. These galaxies are selected  to be [SII]-deficient, massive star forming galaxies (selection criteria further described in \S\ref{sec:sample}). We use COS G140L observations that reach below the Lyman Limit to directly measure the amount of escaping LyC radiation, and to derive the LyC escape fraction. We also study the LyC escape fraction and the nature of correlations with host galaxy properties, and compare the properties of this population of galaxies with the traditional LCEs from the LzLCS+ survey \citep{flury22, flury22b, saldana-lopez22}. Our main goal is to determine whether there exist a real dichotomy in the properties of the leaky galaxies in these two classes, and how this may depend on the physical condition of the host galaxy and its ISM. 

The structure of the paper is as follows. In \S\ref{sec:sample}, we summarize the sample selection for this study. \S\ref{sec:data}  discusses the data acquisition, reduction, and analysis techniques. In \S\ref{sec:results}, we present the results and assess their significance in the context of the existing literature for Lyman continuum emitters. Finally, we summarize our conclusions in \S\ref{sec:conclusion}.

\section{Sample Selection} \label{sec:sample}

\begin{figure*}
    \centering
    \includegraphics[width=\textwidth]{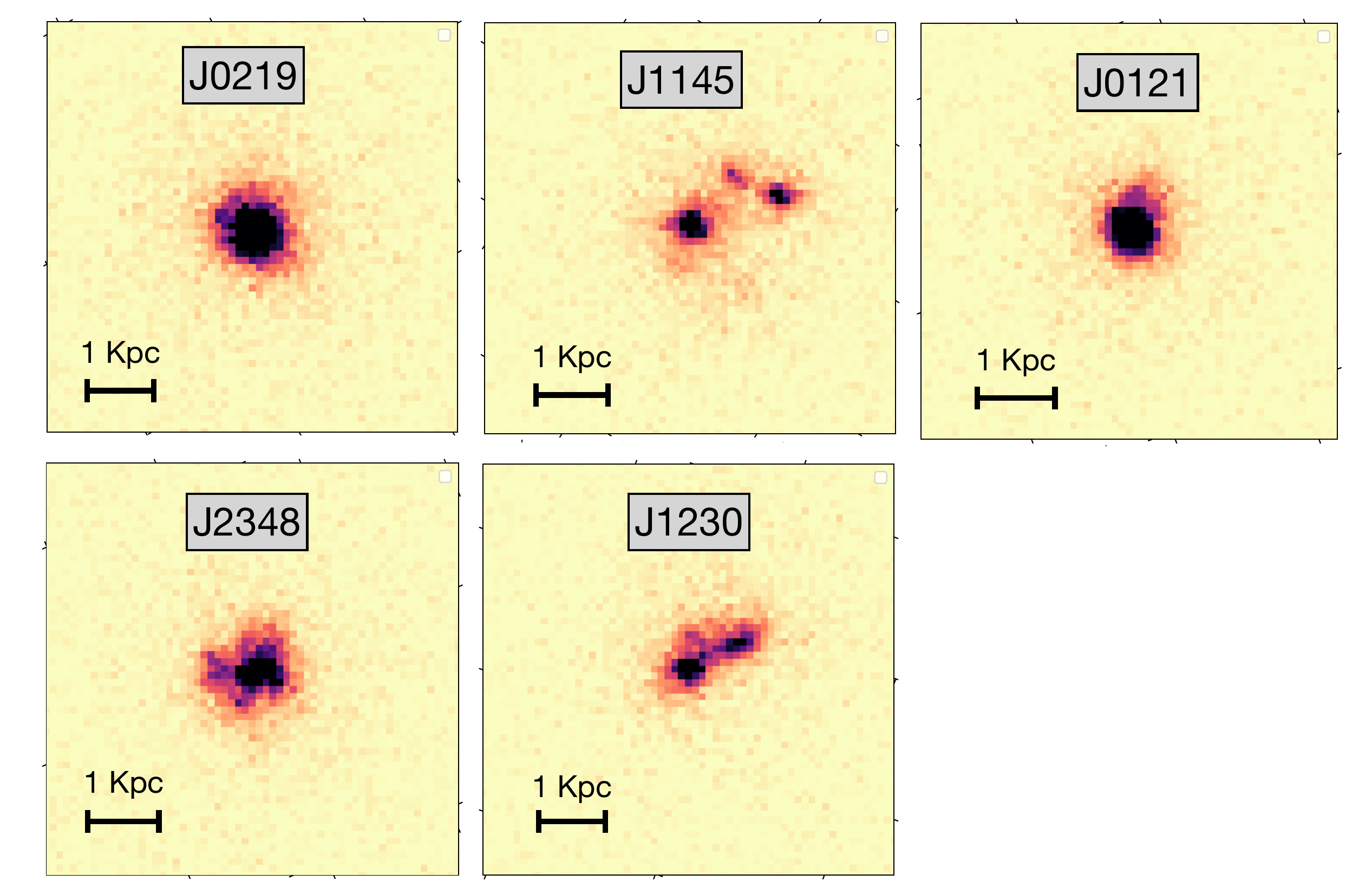}
    \caption{COS near-UV acquisition images of the five [SII]-weak selected targets in this work: (a) J0219 (nonemitter), (b) J1145 (LCE), (c) J0121 (LCE), (d) J2348 (nonemitter), and (e) J1230 (LCE). The images are 1.2$''$ by 1.2$''$ in angular scale. The bar in the bottom left indicates a physical scale of 1 Kpc at the target's redshift.  }
    \label{fig:image}
\end{figure*}

In this study, we investigate a new class of LyC leaky galaxies, selected based on the relatively weak emission of [SII] 6717 \AA, 6731 \AA\  lines compared to those typically observed in star-forming galaxies. Additionally, these galaxies are characterized by compact host galaxies with high stellar masses. This selection criterion is designed to expand the sample of massive leaky starburst galaxies, as previously identified by \cite{borthakur14} and \cite{wang19}. 
Thus, as part of the HST program GO 17220 (PI: T. Heckman), we observed a sample of five galaxies selected from SDSS DR12, WISE near-IR photometry, and GALEX GR6 catalogs based on the following criteria:

\begin{enumerate}
    \item     The SDSS optical spectrum is dominated by starburst activity, showing no indication of an AGN. This is based on the line ratios aligning with ionization from young, hot stars, as indicated by the standard Baldwin, Phillips \& Terlevich (BPT) diagnostic diagram \citep{baldwin81}. There is an absence of emission lines commonly associated with AGN activity, such as [NeV] and He II, and the Balmer line profiles do not exhibit broad wings.

    \item An estimated far-UV flux inside the COS aperture $> \rm 10^{-16} \ erg \ cm^{-2} \ s^{-1} \ $\AA$^{-1} $ at rest frame 1000 \AA. 

    \item An [SII]-deficiency (or $\Delta \rm [SII]$) of at least 0.2 dex relative to the normal star-forming galaxies, defined by the distance from the star formation ridge \citep{kewley01} in the [SII]-BPT. A detailed explanation of the method of calculation of  $\Delta \rm [SII]$ is given in \S\ref{subsec:ancillary}.  As shown by \cite{wang19}, the relative weakness of the [SII] emission lines is a strong indirect tracer of the leakage of Lyman-continuum emission. 

    \item Redshifts higher than 0.25. This ensures that the Lyman Continuum region falls at wavelengths over which the COS has high sensitivity ($>$1150 \AA). 

    \item A seeing-de-convolved half-light radius of $<$0.5$''$ (typically $<$ 2 kpc) based on SDSS u-band images. This mimics the small sizes of galaxies in the epoch of reionization, and ensures that the majority of light is captured within the COS aperture.

    \item Stellar mass $> \rm 10^{10} \ M_{\odot}$ to select massive galaxies \citep{thomas13}.

\end{enumerate}

    The COS NUV acquisition images of the sources are shown in Fig.~\ref{fig:image}. The targets and their relevant properties are listed in Table.~\ref{tab:fesc} and \ref{tab:prop}.  All the HST data used in this work can be downloaded from MAST using the DOI: 10.17909/63mb-ag19.

\begin{table*}
\renewcommand{\arraystretch}{0.9} 
\begin{threeparttable}
	\centering
	\caption{The Lyman Continuum flux and escape fraction of the five massive leakers presented in this work. }
	\label{tab:fesc}
	\begin{tabular}{lcccccccc} 
		\toprule

		ID & z & $\rm F^{obs}_{\lambda LyC} $ & Signif. & $P(>N \vert B)$ & $\rm \frac{F_{900}}{F_{1100}}$ &  $\rm f_{esc, HI}$ & $\rm f_{esc, tot}$ & $\rm f_{esc, dust}$ \\
         &  & [$\rm 10^{-17} \ erg/cm^{2}/s$/\AA] &   &  &   & \\
		\midrule
		J0121 & 0.264 & 2.9$\pm$0.9 & 2.6 &  $4.451\times10^{-3}$ & 0.17$\pm$0.05 & 0.713$\pm$0.23 & 0.0043$\pm$0.0013 & 0.0059$\pm$0.0028 \\
		J0219* & 0.327 & 0.48$\pm$0.93 & $< 2$ & $1.144\times 10^{-1}$& 0.017$\pm$0.03 & 0.05$\pm$0.1 & 0.0035$\pm$0.007 & 0.07$\pm$0.198 \\
		J1145 &  0.287 & 5.38$\pm$1.10 & 5.7 & $4.916\times10^{-9}$ & 0.221$\pm$0.04 & 0.841$\pm$0.16  & 0.0264$\pm$0.0052  &0.0314$\pm$0.0087	\\
		J1230 & 0.342 & 1.58$\pm$0.51 & 2.8 & $2.271\times 10^{-3}$ & 0.078$\pm$0.02 & 0.333$\pm$0.11 & 0.011$\pm$0.004 & 0.03303$\pm$ 0.0169 \\
		J2348* & 0.313 & 0.30$\pm$0.82 & $<$ 2 & $1.820\times 10^{-1}$ & 0.022$\pm$0.06 & 0.14$\pm$0.30 &  0.0025$\pm$0.005 & 0.0178$\pm$ 0.0505 \\
		\bottomrule
  	\end{tabular}
  \begin{tablenotes}
      \small
      \item * marked sources represent Lyman Continuum non emitters and the reported flux and escape fractions are upper limits, derived as described in \S\ref{subsec:result_fesc}.
    \end{tablenotes}
\end{threeparttable}
\end{table*}

Throughout the study, we use the term "massive leakers" to refer to the combined sample of five sources from this work, the three massive, [SII]-deficit LyC leakers previously studied in \cite{wang19} and the massive LyC leaker from \cite{borthakur14}. We compare these massive leaker population with the more traditional LCE sample from the low-z Lyman continuum leaker survey \citep[LzLCS+: ][]{flury22, saldana-lopez22}. The LzLCS+ sample includes  
66 candidate low redshift (z$\sim$ 0.3-0.4) leakers selected from SDSS \& GALEX data archive. The sample spans a broad range in host galaxy parameters like [OIII]/[OII], star formation rate surface density ($\rm \Sigma_{SFR}$),  stellar mass, metallicity, UV slope and others. The program was carried on with  134 orbits of HST/ COS spectroscopy a part of GO 15626 (Cycle 26, PI: A. Jaskot). All the measurements of host galaxy properties of the LzLCS+ sample are obtained from the catalog presented in \cite{flury22, flury22b}. In addition to the 66  galaxies, the catalog also compiles the measurements of 23 other LCE candidates from the literature observed by HST/COS as part of other programs \citep{izotov16, izotov16b, izotov18, izotov18b, izotov21}. The HST/COS observations were reprocessed and the LyC fluxes were recomputed systematically by \cite{flury22} to obtain consistent LyC measurements. We will refer to this combined sample of 89 galaxies as our  LzLCS+ sample.


\section{Data Acquisition and Analysis} \label{sec:data}

\subsection{Processing HST/COS observations} \label{subsec:reduction}

The 12 orbits of HST COS spectroscopy utilized in this study were obtained through the observing program GO 17220 (PI Timothy Heckman).  The G140L grating was used at 800 \AA\ in COS Lifetime Position 4, covering an observed wavelength range of 800-1950 \AA. At the median redshift of our sample, this corresponds to a rest-frame wavelength range of approximately 615-1500 \AA.
 The spectral resolution of the obtained spectrum R $\sim$ 1050 at 1100 \AA. The COS spectroscopic aperture has a field of view of diameter 2.5$''$ and is centered based on the NUV acquisition images taken as part of the program. 

We retrieved the COS data from the MAST archive and
processed the raw spectra using a combination of standard (\texttt{CALCOS}\footnote{\href{https://github.com/spacetelescope/calcos}{https://github.com/spacetelescope/calcos}}) and custom software 
\citep[\texttt{FAINTCOS} \footnote{\href{https://github.com/kimakan/FaintCOS}{https://github.com/kimakan/FaintCOS}}; ][]{worseck16, makan21} to best model the background and optimize the measurement of the LyC. We reduced the data following the method described in \cite{flury22}. The major component of the data reduction is accurately subtracting the dark counts and the background signal, which can contribute significantly to the flux in the LyC region. So we take a few additional steps to mitigate the challenge. We describe them briefly below.

The charge produced by the amplifying microchannel plate in the COS detector is measured by the ``pulse height amplitude" (PHA). The PHA values are triggered by science events, and also spuriously by dark current and geomagnetic activity. This often extends the PHA distribution well beyond the possible values of science events \citep{worseck16, izotov16b}. So, following \cite{flury22}, we screen the PHA values and include only values within the permitted 1-12 range for lifetime position 4, to mitigate dark current and other background effects without excluding
science events.

\begin{figure*}
    \centering
    \includegraphics[width=\textwidth]{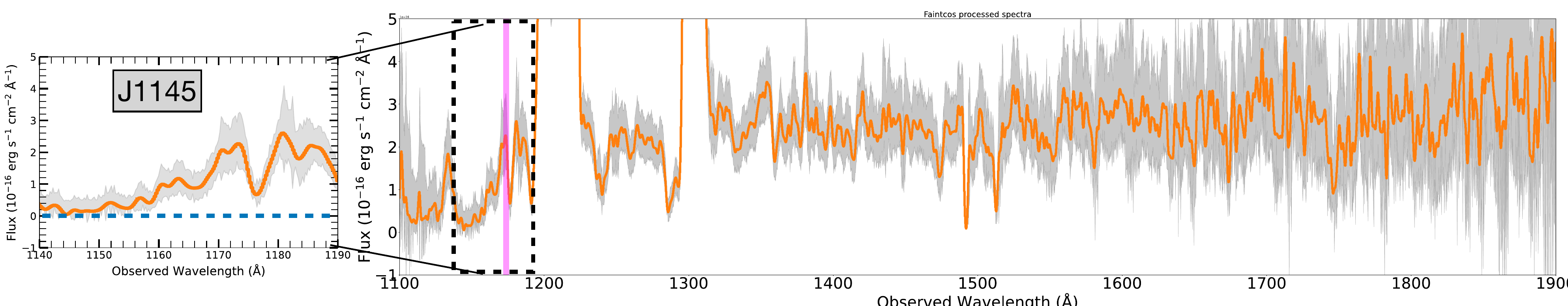}
    \includegraphics[width=\textwidth]{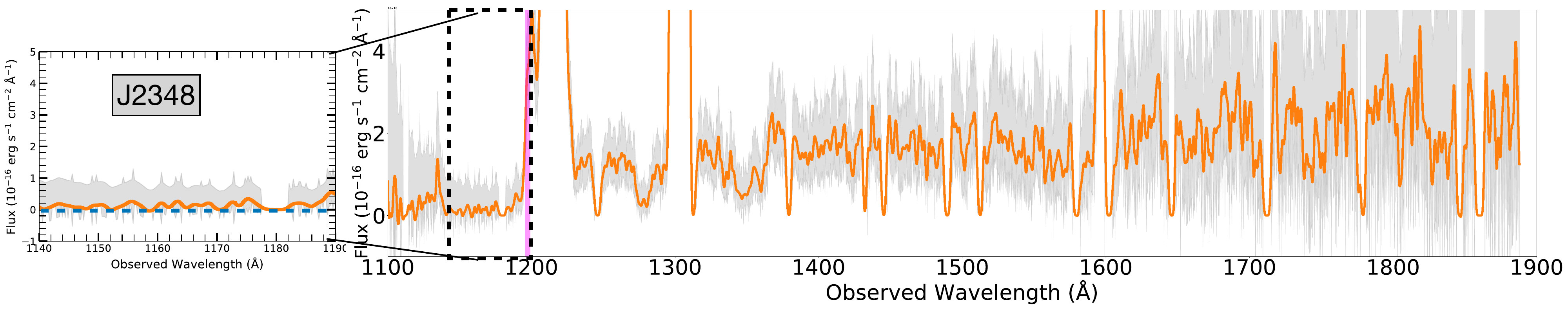}
    \includegraphics[width=\textwidth]{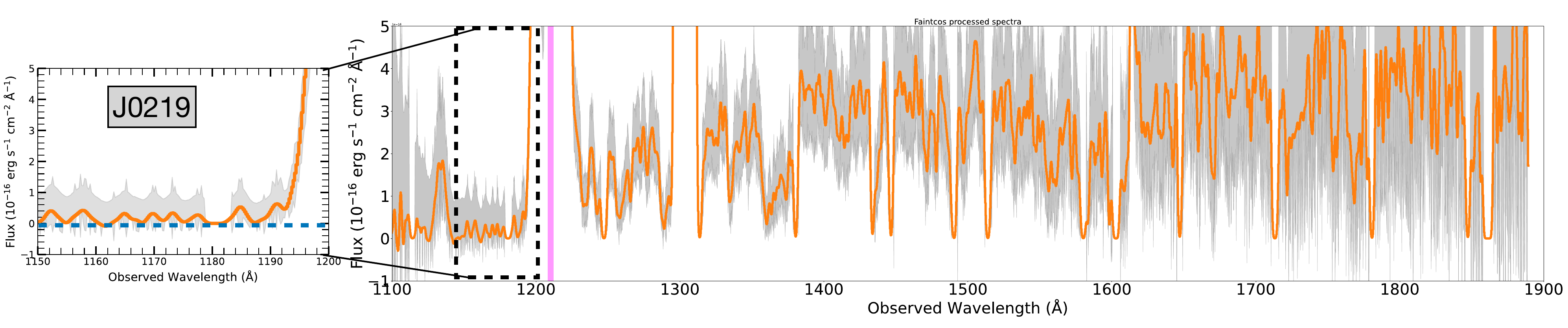}
    \includegraphics[width=\textwidth]{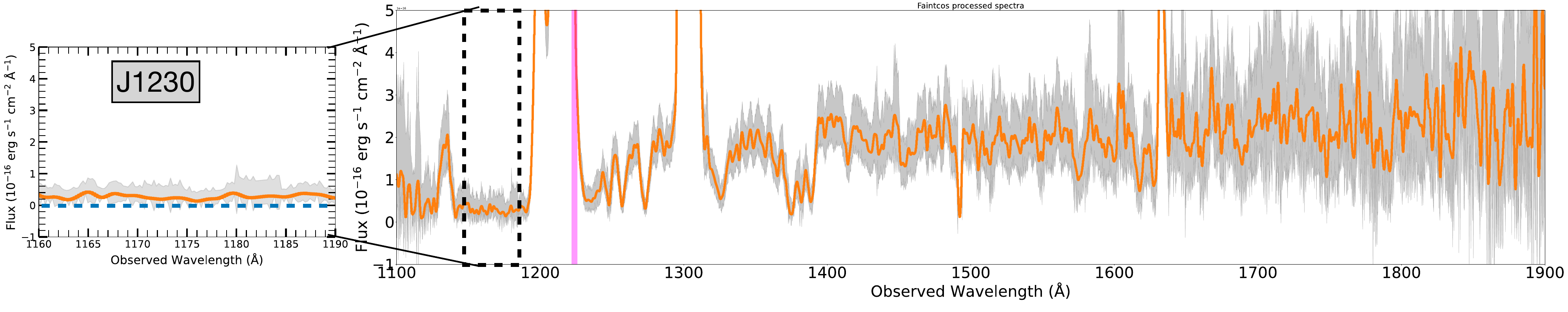}
    \includegraphics[width=\textwidth]{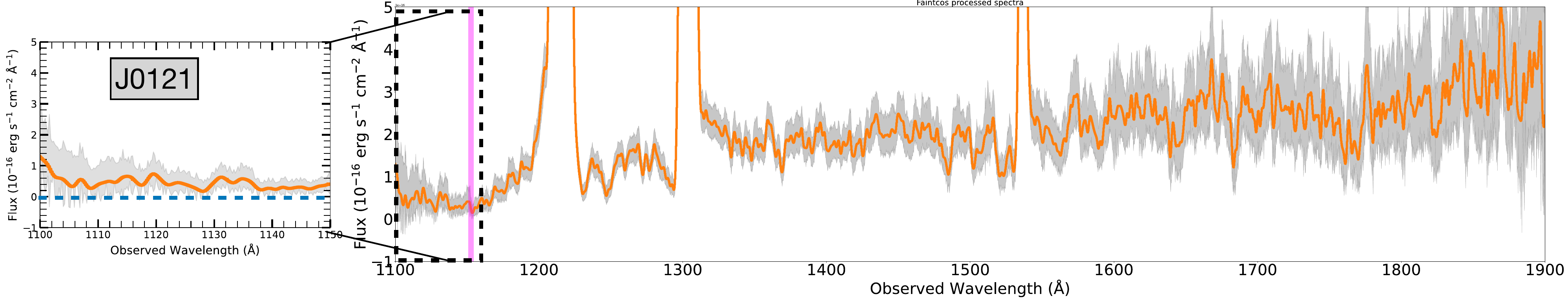}
    \caption{HST/COS spectra (orange) of the five sources analyzed in this work. The left panels show the spectral region around the LyC edge, while the right panels show the full spectra. The gray shaded region indicates the 1$\sigma$ uncertainty in flux density. The location of the Lyman limit at the redshift of the source is indicated by the vertical magenta line. The blue dashed lines in the left panel show the zero flux level. Geocoronal emission from telluric Ly$\alpha$ is clipped in the spectra. }
    \label{fig:spectra}
\end{figure*}

We then process the data through the standard COS pipeline \texttt{Calcos}, version 3.4.7. CALCOS performs the standard reduction steps including flat-fielding, deadtime corrections, wavelength, and flux calibrations. We turn off the native background correction in \texttt{Calcos} and instead subtract a custom super-dark image following the procedure in \cite{leitherer16, wang19, flury22}. A unique super-dark image is created for every science target by combining all the HST/COS dark frames taken within $\pm$1 month of the observing time of the target and computing the average. The $\pm$1 month window is taken to minimize the effects of temporal fluctuations of the dark current. 
We use the custom software package \texttt{FAINTCOS} to estimate the dark current model from the chosen frames and implement the background subtraction to the science spectra. The package computes, for each observation, the cumulative PHA distribution of the chosen dark current observations and matches them with science background images via the Kolmogorov-Smirnoff test. 
To ensure that the super dark image technique has been successful, we compare the spatial variation of the dark model to the science image background and find that they agree reasonably well.  
\texttt{Faintcos} also estimates the contamination from scattered geocoronal Ly$\alpha$ background from \cite{worseck16} model, coadds individual science exposures for each object, and bins them to
Nyquist-sample the G140L resolution of $\sim$1.1 \AA. The final spectra for our 5 sources, shown in Fig.~\ref{fig:spectra}, are now ready for the next steps of analysis.

\subsection{Measuring Lyman Continuum} \label{subsec:measure_lyc}

We smooth the spectra of all sources with a Gaussian kernel with full width at half maximum of about 1.0 \AA, to achieve the native resolution and boost the relatively low signal-to-noise. 
We measure the LyC flux in each source in a rest frame 20 \AA \ window as close as possible to $\rm \lambda_{rest}$ = 900 \AA, while simultaneously avoiding contamination from Ly$\alpha$ and NI $\lambda$1200 geocoronal emission lines. 
Thus, we avoid wavelengths above $\rm \lambda_{obs}$ = 1180 \AA \ \citep[similar to ][]{flury22}. We also do not choose the spectral range to include wavelengths below $\rm \lambda_{obs}$ = 1100 \AA, since the sensitivity of the G140L grating and the signal-to-noise of the spectra decline rapidly.  Thus, to compute LyC flux ($\rm F_{\lambda LyC}$) for our sources at z = 0.26 $-$ 0.34, we compute the mean background subtracted flux density in the chosen 20 \AA \ spectral window generally within $\rm \lambda_{rest}$ = 860-910 \AA, such that the above conditions are met.  
This ensures a uniform measurement of the LyC flux $\rm F_{ \lambda LyC}$ across our entire sample and is also consistent with the LzLCS+ survey to enable direct comparison. 

To robustly assess the LyC detections, we calculate the probability, $P(>N \mid B)$, that the total observed (gross) counts $N$ within the LyC spectral window arise from the distribution of background counts $B$, following the definition of \cite{worseck16} and the methodology adopted by \citet{flury22}. We employ the Python package \texttt{FeldCous} \citep{feldcous, feldman98} to compute $P(>N \mid B)$ and derive the detection significance, using the same prescription as \cite{flury22}, to identify galaxies that meet the conventional $2\sigma$ detection threshold. The resulting detection significances for our sources are listed in Table~\ref{tab:fesc}.

One of the main goals of this investigation is to explore the roles of both dust extinction and absorption by HI in determining the fraction of escaping LyC photons. We also want a mix of indicators that includes a simple model-independent parameter. We will therefore characterize the escape of LyC photons by reporting three different quantities: 

\begin{itemize}
    \item The direct flux ratio $\rm F_{ \lambda LyC}$/F$_{\rm \lambda 1100}$ (often labelled as $\rm F_{900}/F_{1100}$) which we denote as Method 1.
    \item $\rm f_{esc, tot}$, which is  the ratio of F$_{\rm \lambda LyC}$ to the intrinsic contribution from the stars, as inferred from fitting models of the observed UV continuum (often labelled as absolute escape fraction $\rm f_{esc, abs}$). This is Method 2.
    \item $\rm f_{esc, corr}$ (often labelled as $\rm f_{esc, HI}$), which is corrected for the effects of dust attenuation and represents the fraction of LyC photons that would escape after surviving neutral hydrogen absorption, assuming no dust is present. This is denoted by Method 3.
\end{itemize}

To derive $\rm F_{ \lambda LyC}$/F$_{\rm \lambda 1100}$, we transform the observed spectra to the rest frame of the galaxy using SDSS spectroscopic redshifts. We compute $\rm F_{ \lambda LyC}$ by calculating the mean flux density in the LyC spectral window of width 20 \AA, as discussed above. F$_{\rm \lambda 1100}$ is calculated by taking average flux in 1090 $ < \rm \lambda_{rest} < $ 1100 \AA, i.e. keeping the spectral window width to be the same as $\rm F_{ \lambda LyC}$ for consistency. Our method is consistent with previous studies, allowing for direct comparison \citep{wang19, flury22}.

We estimate $\rm f_{esc, tot}$ by taking the ratio of the observed flux in the LyC region and the intrinsic flux in the same wavelength range derived from the best-fit stellar model spectra. The stellar continuum modeling is performed using a customized PYTHON script \texttt{FICUS} \footnote{The FICUS code is publicly available here: \href{https://github.com/asalda/FiCUS.git}{https://github.com/asalda/FiCUS.git}}, which stands for \textit{FItting the stellar Continuum of UV Spectra} \citep{saldana-lopez22, chisholm19}. The continuum modeling is achieved in FICUS by fitting every observed spectrum with a linear combination of multiple bursts of single-age and single-metallicity stellar population models and performing a chi-squared minimization. By default, FICUS inputs the fully theoretical STARBURST99 single-star models without stellar rotation \citep[SB99; ][]{leitherer11, leitherer14} using the Geneva evolution models \citep{meynet94}, and computed with the WM-BASIC method \citep{pauldrach01, leitherer10}. The SB99 models assume a \cite{kroupa01} initial mass function (IMF) with a high-(low-)mass exponent of 2.3 (1.3), and a high-mass cutoff at 100 $\rm M_{\odot}$. The spectral resolution of the SB99 models is R$\sim$ 2500, and it remains approximately constant at FUV wavelengths. Four different metallicities (0.05, 0.2, 0.4, and 1 $\rm Z_{\odot}$) and ten
ages for each metallicity (1, 2, 3, 4, 5, 8, 10, 15, 20, and 40 Myr)
were chosen as a representative set of 40 models for our UV spectra. Finally, a nebular continuum was generated by self-consistently
processing the stellar population synthesis models through the
CLOUDY V17.0 code 4 \citep{ferland17}, assuming same gasphase
and stellar metallicities, an ionization parameter of log (U)
= -2.5, and a volume hydrogen density of $\rm n_{H}$ = 100 $\rm cm^{-3}$ \citep{chisholm19}.

\begin{figure*}
    \centering
    \includegraphics[width=\textwidth]{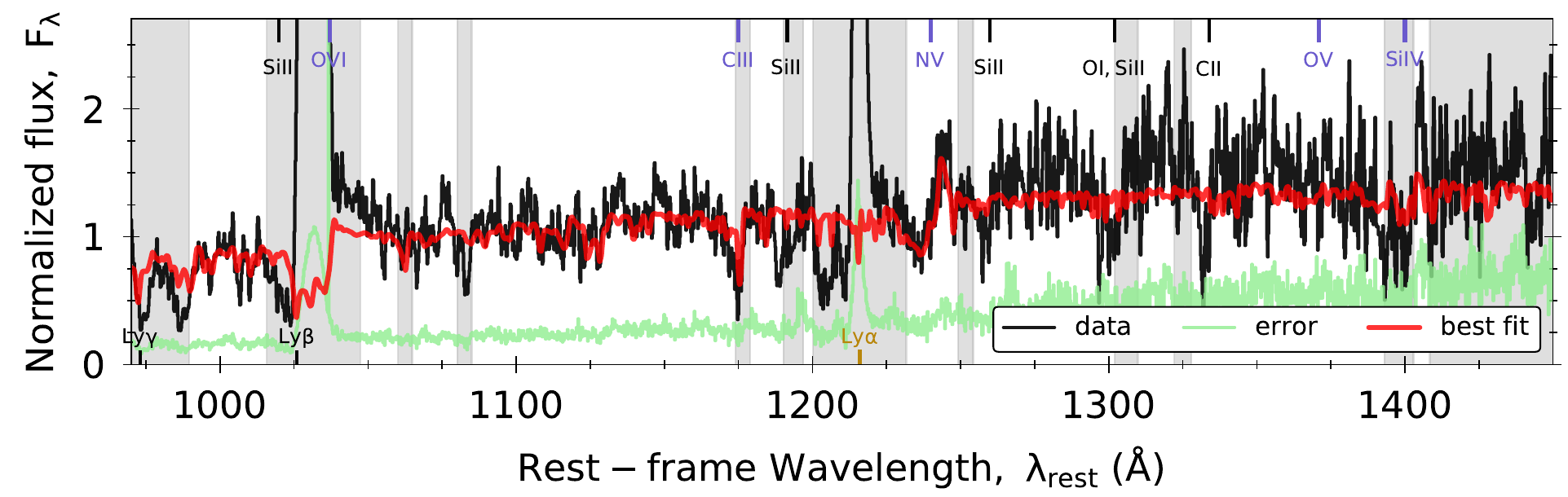}
    \caption{SED fitting results for galaxy J0121 with FICUS. The observed and error spectra are displayed in black and light green. The
best-fitting stellar continuum is overplotted in red, while the spectral regions masked during the fit are shown in grey. The spectrum, shown in $\rm F_{\lambda}$ units, has been normalized over 1350–1370 \AA. The most prominent stellar features, nebular and ISM absorption lines are indicated with black vertical lines at
the top part of the figure.  }
    \label{fig:stellar-model-fit}
\end{figure*}

To perform the fitting, the observed spectra are first manually placed into the rest-frame by multiplying by the corresponding 1/(1 + z) factor. Both the spectra and the models are then normalized by the median flux within a
wavelength interval free of stellar and ISM features (1350–1370 \AA),
and all the fits are performed in the same rest-wavelength range
(930–1425 \AA). Finally, the models
are convolved by a Gaussian kernel to the instrumental resolution
(R $\sim$ 1000). The best-fit stellar model spectrum for each source is integrated over the optimal 20 \AA \ wide LyC spectral window to estimate the modeled (i.e. intrinsic) LyC flux ($\rm F_{\lambda LyC, int}$).  Fig.~\ref{fig:stellar-model-fit} shows the FICUS fitting for one of the galaxies 
(ID: J0121, at z = 0.264), with the observed spectrum in black
and fitted stellar continuum in red.

We use FICUS for stellar continuum modeling to be consistent with the results from the LzLCS+ study. However, for one of our sources (ID: J1145), the FICUS fit fails due to inadequate signal to noise, particularly at the red end of the spectrum. So for J1145, we directly generate a grid of synthetic spectra from STARBURST99 models, and assume a continuous and constant star formation rate (SFR) = 1 $\rm M_{\odot}$, and a Kroupa Initial mass function \citep{kroupa01}. We generate a total of eight sets of SB99 models based on two choices each for burst age ($\rm 10^7$ and $\rm 10^8$ yr), metallicity (solar or 1/7 solar), and whether or not stellar rotation is present. We assume the stellar population evolves from the zero-age main sequence using the evolutionary models of the Geneva Group (similar to the FICUS assumptions).
 The model spectra are interpolated into the same wavelength array as its corresponding COS spectrum,  convolved with the same Gaussian kernel, and normalized to the common flux level (1350–1370 \AA) -- following the same procedure as the rest of the sources. The best fit is chosen by eye; more specifically, we closely examine the match of the two strong stellar wind features -- O VI 1032,1038 and N V 1238,1242 \citep[similar to 
 ][]{wang19}. These P-Cygni features trace the most massive stars that are responsible for producing most of the ionizing continuum. We multiply the SB99 model-derived LyC flux with the total SFR since the stellar models were assumed to have a constant SFR = 1$\rm M_{\odot}/yr$ in the model grid. The SFR was derived using WISE band 3 and band 4 data  to estimate a rest-frame 24 micron luminosity and then using the relationship in \cite{kennicutt12} to get SFR$_{\rm IR}$. We then use this SFR and the SB99 model to obtain the intrinsic LyC flux ($\rm F_{\lambda LyC, int}$). 

Then $\rm f_{esc, tot}$ is essentially the ratio of $\rm F_{ \lambda LyC}$/ $\rm F_{\lambda LyC, int}$. The measured value of this escape fraction depends on both the effects of dust extinction and photoelectric absorption of the LyC due to hydrogen. The computed $\rm f_{esc, tot}$ values derived from the \texttt{FICUS} package and those obtained directly from Starburst99 differ by less than 1\% for all our sources (except J1145 where the FICUS fit fails), confirming a robust characterization of the escape fraction independent of the method used.

For method 3, 
we measure the dust-corrected escape fraction $\rm f_{esc, corr}$ which is the ratio of emergent
LyC flux to the intrinsic flux, after removing the effects of dust extinction. To determine the amount of dust extinction, we first measure the rest-frame 1500 \AA \ flux and convert that to luminosity. We use the $\rm L_{1500}$ to  estimate  $\rm SFR_{UV}$ using the \cite{kennicutt12} relation, and derive the attenuation at 1500 \AA \ by:
\begin{equation}
    \rm A_{1500} = 2.5 \times log_{10} (SFR_{IR}/SFR_{UV})
\end{equation}

According to the far-UV dust attenuation law by \cite{reddy16}, E($\lambda_1$-$\lambda_2$) is given by 0.226 $\times$ A$_{1500}$, where $\lambda_1$ = 900 \AA\ and $\lambda_2$ = 1100 \AA, based on an extrapolation of their fit to data between 950 and 2000 \AA. The dust-corrected ratio $\rm (F_{\lambda LyC}/F_{\lambda 1100})_{corr}$ is then calculated as $\rm 10^{0.4 \times 0.226 \times A_{1500}} \times (F_{\lambda LyC}/F_{\lambda 1100})$. The final \fesc\ estimate is obtained by dividing the dust-corrected $\rm (F_{\lambda LyC}/F_{\lambda 1100})_{corr}$ by the intrinsic $\rm (F_{\lambda LyC}/F_{\lambda 1100})_{int}$ derived from the modeled spectrum. This $\rm f_{esc, corr}$ value measures only the photoelectric absorption of LyC photons by HI in the absence of dust. Thus throughout the text, $\rm f_{esc, corr}$ is also referred to as $\rm f_{esc, HI}$. Table \ref{tab:fesc} presents the Lyman Continuum  escape fractions, along with their associated $1\sigma$ uncertainties, calculated using all three methodologies discussed above for our five target objects. 


    
\begin{table*} \setlength{\tabcolsep}{3pt}
\renewcommand{\arraystretch}{0.8} 
	\centering
	\caption{The host galaxy properties of the five massive leakers presented in this work. }
	\label{tab:prop}
	\begin{tabular}{lccccccccc} 
		\toprule
		ID &   [OIII]/[OII] & H$\beta$ EW & Ly$\alpha$ EW & $\rm E(B-V)_{UV}$ & 12 + log (O/H) & $\rm Log_{10} M_{\star}$ & $\rm SFR_{UV}$ & $\rm \Sigma_{SFR, UV} $ & $\rm \Delta[SII]$ \\
         &     & [\AA] & [\AA] &  &  & [$\rm M_{\odot}$] & [$\rm M_{\odot} \ yr^{-1}$] & [$\rm M_{\odot} \ yr^{-1} \ kpc^{-2}]  $ &   \\
		\midrule
		J0121 &   0.515$\pm$0.016 &	21.1 $\pm$0.4 & 43.1$\pm$3.5 & 0.298$\pm$0.007 &8.59$\pm$0.13 & 10.51$\pm$0.1 & 60$\pm$2 & 130.9$\pm$5.2 & -0.278 $\pm$ 0.08 \\
		J0219 &  0.52$\pm$0.019 & 	13.0$\pm$0.5 & 8.0$\pm$2.1 & 0.235$\pm$0.013 & 8.48$\pm$0.06 & 10.5$\pm$0.11 & 26$\pm$4 &22.1$\pm$4.1 & -0.457 $\pm$ 0.22\\
		J1145 &   0.490$\pm$0.015 & 25.9$\pm$0.8 &  1$\pm$5 & 0.345$\pm$0.009 & 8.48$\pm$0.07 & 10.00$\pm$0.03 & 25$\pm$1 & 6.7$\pm$0.7 & -0.249 $\pm$ 0.05\\
		J1230 & 0.437$\pm$0.018 & 18.0$\pm$0.6 & 11.2$\pm$2.5 & 0.300$\pm$0.015 & 8.49$\pm$0.12 & 10.40$\pm$0.03 & 36$\pm$2 & 13.5$\pm$2.5 & -0.213 $\pm$ 0.11\\
		J2348 &   0.607$\pm$0.02 & 16.2$\pm$0.5 & 32.0$\pm$7.5 & 0.351$\pm$0.01 & 8.53$\pm$0.13 & 10.60$\pm$0.02 & 53$\pm$2 & 31.4$\pm$3.4 & -0.169 $\pm$ 0.03\\
		\bottomrule
	\end{tabular}
\end{table*}

\subsection{Measurement of ancillary parameters} \label{subsec:ancillary}

This paper aims to study the LyC escape fraction and its dependence on host galaxy properties compared with the LzLCS+ survey. Here we list the ancillary parameters we measure. 

We measure the equivalent width (EW) of the Ly$\alpha$ emission line from the extracted COS spectra. The continuum is modeled by fitting a linear function to the continuum within a spectral window of $\pm$ 30\AA \ on either side of the emission line using an iterative sigma clipping technique to exclude noise spikes. To measure \lya \ EW, the observed spectrum is normalized by dividing the spectrum by the modeled continuum and then corrected for redshift.  
The fluxes and EWs of the rest-optical emission lines are taken from Sloan Digital Sky Survey (SDSS) spectra, as listed in the value-added catalog produced by the Portsmouth group \citep{thomas13}.
 A signal-to-noise ratio (S/N) cut of 5 is imposed for any emission line detection. The flux values for [OIII]5007 and [OII]3726, 3729\AA \ emission lines are obtained from the catalog to compute the [OIII]/[OII] ratio.

These are dusty systems, so our primary measure of SFR is based on both $\rm SFR_{IR}$ from the WISE mid-IR photometry, and $\rm SFR_{UV}$ based on the observed continuum flux measured at 1500\AA \ (see above). We do not use a SFR based on the Balmer emission-lines since these can significantly underestimate the SFR in cases in which only a fraction of the ionizing photons are absorbed by HI in the galaxy ISM.



We measure the [S II] deficiency ($\rm \Delta [SII]$) in a differential sense: as a
quantity relative to the majority of star-forming galaxies in the
SDSS survey. To do this, we first construct the BPT/VO [S II]/H$\alpha$ versus [O III] 5007/H$\beta$ diagnostic diagram \citep{baldwin81, veilleux87} for our objects using the flux values from the Portsmouth catalog \citep{thomas13}. We then subtract the measured [SII]/H$\alpha$ from the mean BPT/VO relation derived for a large sample of SDSS star-forming galaxies from \cite{wang19}. Thus, we define the [SII] deficiency as a galaxy's displacement in log([SII]/H$\alpha$) from the \cite{wang19} polynomial fit to the mean star-forming relation, as follows:

\begin{multline*}
     y = -0.487 + 0.014 \xi + 0.028 \xi^{2} -0.785 \xi^{3}\\
    - 3.870 \xi^{4} + 0.446 \xi^{5} + 8.696 \xi^{6} +0.302 \xi^{7} -6.623 \xi^{8} 
\end{multline*}

where $\xi$ is log([OIII]/H$\beta$) and y is log([SII]/H$\alpha$). Thus, $\Delta$[SII] =  log([SII]/H$\alpha$)$_{\rm observed}$ - y

We use the COS near-UV images to compute the half-light radius for a given galaxy. All targets are located well within the COS aperture radius of 1.2$''$. We accounted for the vignetting of the COS aperture, where the effective radius of the aperture diminishes to $\sim $ 0.4 $''$, while calculating the half light radius. 
We estimate the background signal from the mean of an annulus of $\rm r_{in} = 0.9''$ and $\rm r_{out} = 1.1''$. The half-light radius is calculated by determining a circle that encloses half of the total near-UV emitted light in the background subtracted image within the 0.4 $''$ vignetted aperture. 


Finally, the oxygen abundance of the interstellar medium (ISM) is calculated from the strong emission line tracers following \cite{pettini04}:

\begin{equation}
    \rm 12 + log_{10} (O/H) = 8.73 - 0.32 \times O3N2
\end{equation}

where $\rm O3N2 = log_{10}\frac{[OIII]\lambda 5007/H\beta}{[NII]\lambda 6584/ H\alpha}$. 
The relation is valid for -1 $<$ O3N3 $<$ 1.9 and is insensitive to dust extinction. Note, for the LzLCS+ sample, the oxygen abundance is calculated form the [OIII]$\lambda$4363\AA \ line. We provide these measurements of the host galaxy properties of our sources in Table.~\ref{tab:prop}.

\section{Results \& Discussions} \label{sec:results}

\subsection{LyC escape fraction } \label{subsec:result_fesc}

\begin{figure*}
    \centering
    \includegraphics[width=\textwidth]{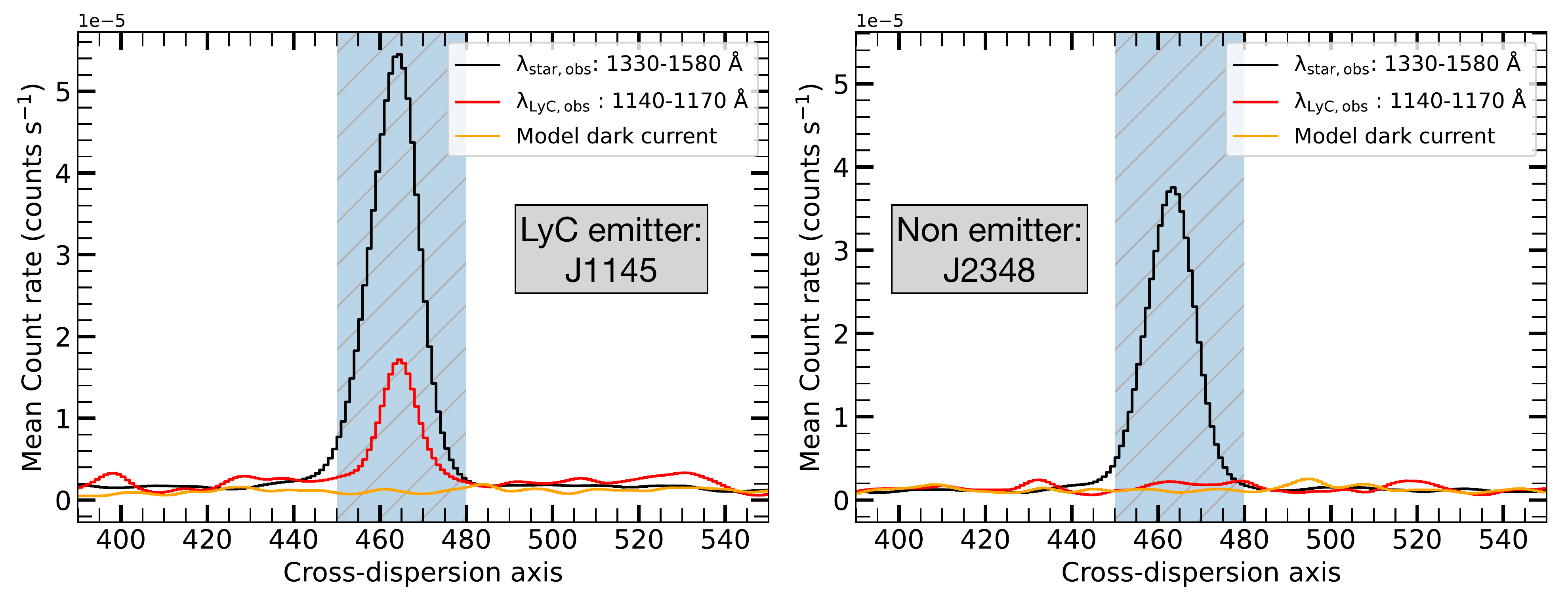}
    \caption{Example spatial cross section of the two-dimensional spectrum for a well-detected Lyman Continuum Emitter with $>3\sigma $ detection (left) and a nonemitter (right) from this work. The black line represents the nonionizing starlight continuum, the yellow line represents the model dark current, red line represents the LyC counts, and the blue hatched stripe represents the spectral extraction aperture.}
    \label{fig:cross-dispersion}
\end{figure*}

We detect a significant ($\geq 2\sigma$) flux in the LyC region below the Lyman edge for 3 out of our 5 galaxies. For these three detected sources, we report the mean background-subtracted flux density in the LyC spectral window (as explained in \S \ref{subsec:measure_lyc}). In case of the two non-detections, we take the 84th percentile of the background distribution to be the upper limit on $\rm F^{obs}_{\lambda LyC}$ following previous LzLCS+ studies \citep{flury22}. 
The corresponding $1\sigma$ uncertainties in the reported LyC fluxes are derived by summing the two error sources in quadrature: the $1\sigma$ errors from the HST/COS spectra within the LyC spectral window and the $1\sigma$ uncertainties from bootstrapped residual dark current fluxes across the entire detector region. To confirm the $>2\sigma$ LyC detections, we also examine the cross-dispersion profiles of the two-dimensional spectra of our sources in the LyC window and compare with the dark current flux levels and the non-ionizing stellar counterpart centered at $\rm \lambda_{rest} \sim $ 1100 \AA. Fig.~\ref{fig:cross-dispersion} shows example profiles for a LyC detected and non-detected source.

\begin{figure}
\vspace{4mm}
    \centering
    \includegraphics[width=0.47\textwidth]{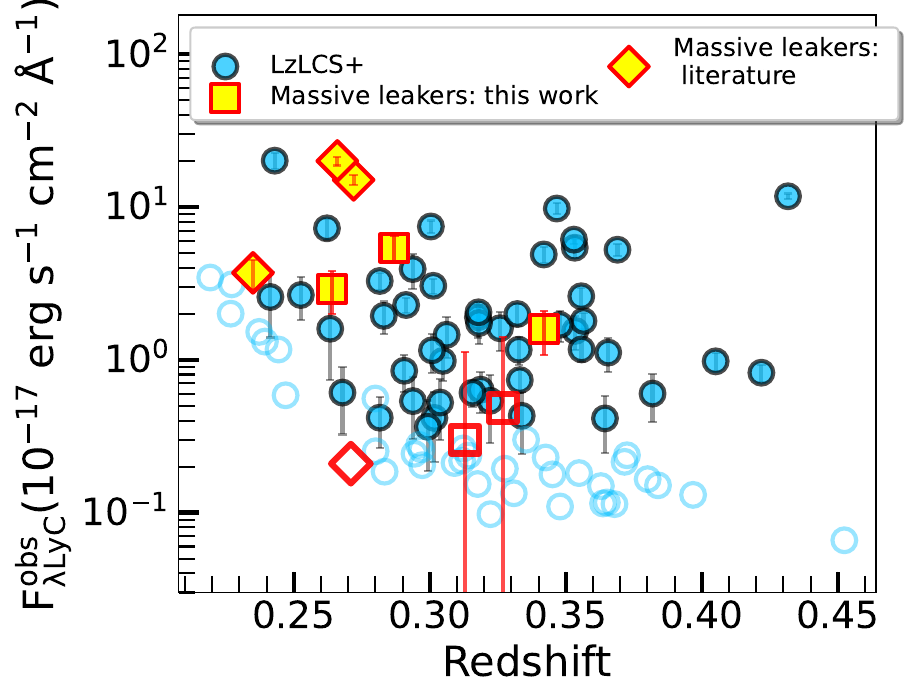}
    \caption{LyC flux  vs. redshift for the massive leaker population from this study (yellow squares) alongside published sample from the  literature \citep[yellow diamonds: ][]{wang19, borthakur14}. Measurements from the LzLCS+ sources \citep{flury22} are represented by blue circles. Flux upper limits for non-emitters are represented by transparent, unfilled symbols.}
    \label{fig:fesc}
\end{figure}

For the LyC detected sources, we report flux densities $\rm F_{\lambda LyC}$ (and $1\sigma$ errors)
of 5.38 ($\pm 1.10)\times \rm 10^{-17} \ erg \ cm^{-2} \ s^{-1}$ \AA$^{-1}$ for J1145, 2.9 ($\pm 0.90) \times \rm 10^{-17} \ erg \ cm^{-2} \ s^{-1}$ \AA$^{-1}$ for J0121, and 1.58 $(\pm 0.51) \times \rm 10^{-17} \ erg \ cm^{-2} \ s^{-1}$ \AA$^{-1}$ for J1230. The detection significances for the three LyC-detected sources, computed using \texttt{FeldCous} (see \S\ref{subsec:measure_lyc}), are listed in Table~\ref{tab:fesc}. For the two non-detected sources that fall below the $2\sigma$ threshold, we report upper limits. Following the classification scheme of \cite{flury22}, two of our LyC leakers (J0121 and J1230) are classified as weak detections ($\sim 2\sigma$), while J1145 qualifies as a strong LyC source with $>5\sigma$ significance.  In Fig.~\ref{fig:fesc}, we present the LyC fluxes as a function of redshift for our massive leaker sample (yellow squares), alongside previously published data on similarly massive LyC leakers from \cite{wang19} and \cite{borthakur14} (yellow diamonds). These are compared to the traditional LzLCS+ sample, as detailed in \cite{flury22b} (see \S\ref{sec:sample} for a comprehensive description of the sample).



As mentioned in \S\ref{subsec:measure_lyc}, we report three different ways to measure the escape fraction.  The first and the simplest method  is computing the ratio of the observed fluxes in the LyC region to that of rest wavelength of 1100 \AA \ $\rm F_{ \lambda LyC}$/F$_{\rm \lambda 1100}$. We denote this as method 1. The rest-frame 1100 \AA \ flux provides a measurement for the non-ionizing stellar continuum, and 
thus, the mean ratio of LyC to 1100 \AA \ 
is an indirect proxy for \fesc. It is very useful since it is directly derived from observed spectrum in a model-independent way without any assumptions about stellar populations or other galaxy properties.

\begin{figure*}
    \centering
    \includegraphics[width=\textwidth]{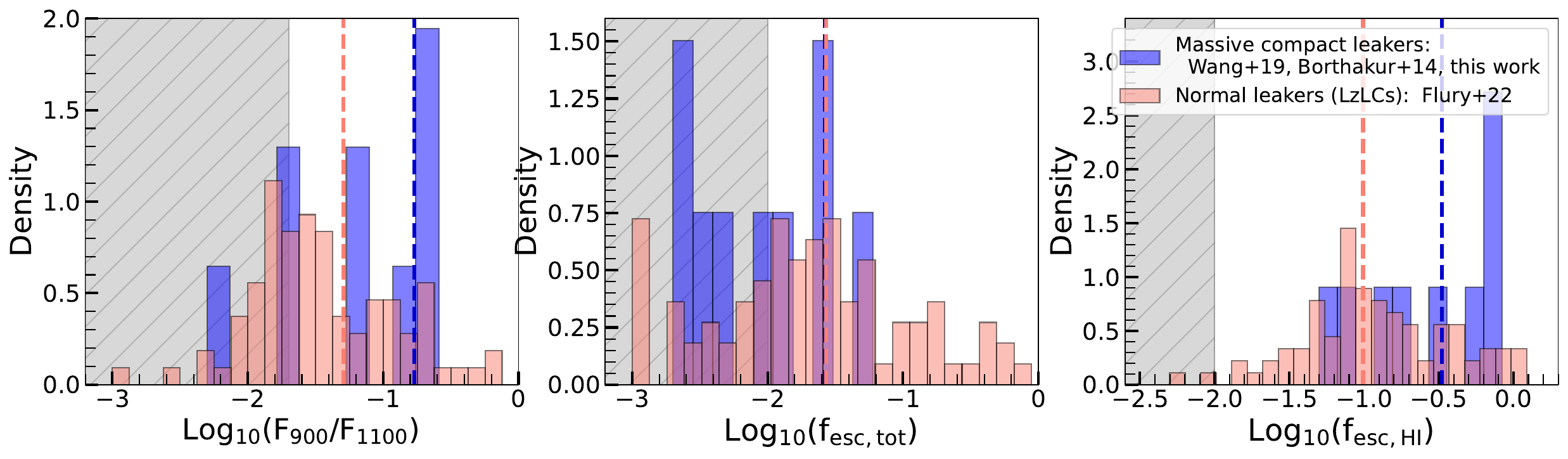}
    \caption{Histogram distributions of three different estimates of LyC escape fraction (as detailed in \S\ref{subsec:measure_lyc}) for the combined massive leaker population from this work, \cite{wang19} and  \cite{borthakur14} (blue), alongside the LzLCS+ sample from \cite{flury22} (red). The left panel displays the $\rm F_{\lambda LyC}/F_{1100}$ values derived from direct flux ratios (method 1), the middle panel shows the distribution of $\rm f_{esc, tot}$ obtained from stellar model fits to the observed UV continuum (method 2), and the right panel presents $\rm f_{esc, HI}$ estimates representing LyC escape due to HI absorption only in the absence of dust (method 3). The mean of the LyC-detected sources are indicated by dashed vertical lines (red: LzLCS+ sample, blue: massive leakers). The gray-shaded hatched region shows the non-emitters with upper limits. }
    \label{fig:histogram}
\end{figure*}

Fig.~\ref{fig:histogram} shows the histogram of \fesc \ measured from three different methods as described in \S\ref{subsec:measure_lyc} in three panels. The left panel shows the distribution of $\rm F_{ \lambda LyC}$/F$_{\rm \lambda 1100}$ for the massive leakers population (blue) and LzLCS+ sample (red).  The mean $\rm F_{ \lambda LyC}$/F$_{\rm \lambda 1100}$ for the massive leakers which are LyC detected = 0.17, compared to a mean value of 0.05 obtained from the LzLCS+ sample. The three LyC detected sources from this study 
exhibit a mean $\rm F_{ \lambda LyC}$/F$_{\rm \lambda 1100}$ to be  0.12, which is a factor of 2 higher than the \cite{flury22} sample.

$\rm F_{ \lambda LyC}$/F$_{\rm \lambda 1100}$ is directly observable, but is  sensitive to a number of factors in addition to the LyC escape fraction, including dust attenuation, SFR, metallicity, and burst ages -- complicating its interpretation. Hence, we computed two other estimates of \fesc \  as described in \S\ref{subsec:measure_lyc}. 
There are two main sources of opacity provided by the galaxy ISM which play a primary role in deciding what fraction of LyC photons can escape into the IGM. The first source of opacity is photoelectric absorption due to neutral Hydrogen, which produces a rapid drop in flux density at the Lyman edge. The second is dust. Dust absorption suppresses the flux by a smoothly varying function that increases with decreasing wavelength, thus making UV flux susceptible to its effects. However, it doesn't produce any sharp change in flux around the Lyman edge.

To consider the escape fraction due to the effect of both photoelectric absorption and the presence of dust, we use method 2, i.e. measuring $\rm f_{esc, tot}$. It involves the ratio between the observed flux in the LyC region to the intrinsic flux from the non-ionizing stellar counterpart derived from stellar population models to provide a measure of the fraction of the LyC photons that suffer both photoelectric absorption by atomic hydrogen as well as dust absorption in the ISM. This yields $\rm f_{esc, tot}$ = 0.0264, 0.0043, and 0.011 for J1145, J0121 and J1230. The estimated escape fractions  are thus rather small. The upper limits for the remaining two non-detected sources are 0.0025 and 0.0035 for J2348 and J0219 respectively. Fig.~\ref{fig:histogram} middle panel shows the comparison of $\rm f_{esc, tot}$ distribution between LzLCS+ and the massive leakers. The detected massive leakers exhibit a mean $\rm f_{esc, tot}$ $\sim$ 0.0264 (shown by blue vertical line) $-$ remarkably similar to the detected LzLCS+ sample (mean $\rm f_{esc, tot} \sim $ 0.027).

\begin{figure}
    \centering
    \includegraphics[width=0.49\textwidth]{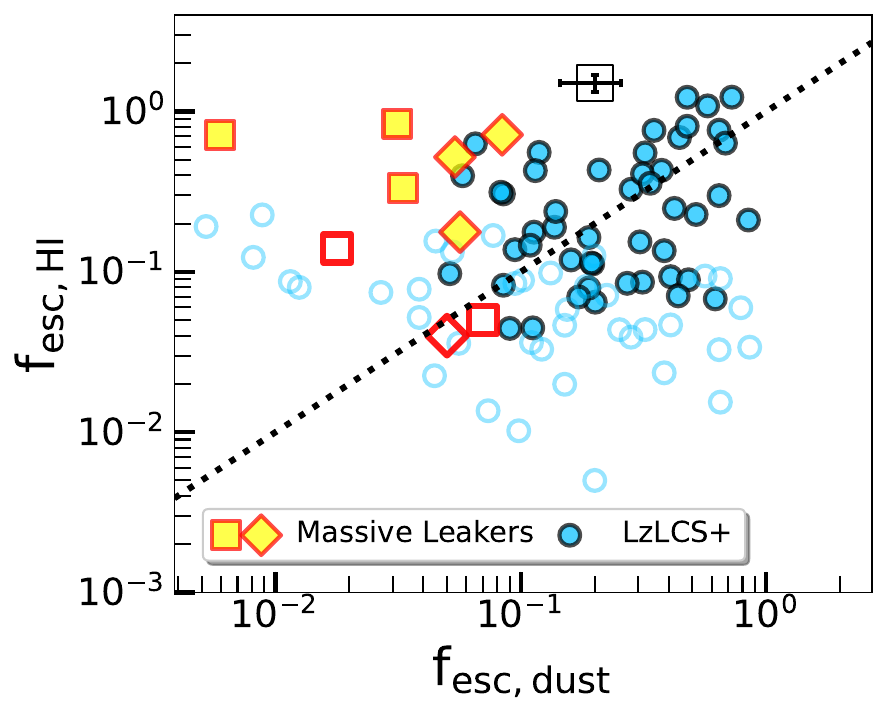}
    \caption{The fraction of LyC photons escaping into the IGM is determined by two main factors: HI absorption and dust attenuation. Here, LyC escape fractions considering only HI absorption $\rm f_{esc, HI}$ are compared with those considering only dust attenuation $\rm f_{esc, dust}$ for the LzLCS+ sample \citep[circles;][]{flury22} and massive leakers \citep[squares: this study, diamonds: literature sample from][]{wang19, borthakur14}. Hollow symbols represent upper limits for non-emitters. Characteristic $1\sigma$ uncertainty of our sample is shown as black errorbar symbol. The dotted line indicates a one-to-one equivalence between the two \fesc\ estimates. Most massive leakers lie above this line, with $\rm f_{esc, HI} > f_{esc, dust}$, suggesting their LyC escape fractions would be higher if not significantly reduced by dust attenuation, unlike the LzLCS+ leakers.
 }
    \label{fig:fesc-HI-dust}
\end{figure}

In the final step (method 3), we correct these estimates for the effect of dust absorption, to derive an estimate of how much of the LyC photons would escape only considering photoelectric absorption by hydrogen. The difference in estimates between method 2 and 3 enables us to isolate the effect of neutral Hydrogen and dust absorption on LyC escape fractions. We employ the method described in \S\ref{subsec:measure_lyc} to determine $\rm f_{esc, HI}$ (or $\rm f_{esc, corr}$), i.e. the \fesc\ determined by photoelectric absorption by hydrogen only.  We obtain $\rm f_{esc, corr}$ = 0.841, 0.583 and 0.363 for  J1145, J0121 and J1230, with a considerably high mean value of 0.59. We calculated $\rm f_{esc, corr}$ for the LzLCS+ sample  using a method consistent with our own analysis and found a mean  $\rm f_{esc, corr}$ of 0.2 for the LzLCS+, which is approximately three times lower than that of the massive leakers identified in this study. This indicates that a large fraction of LyC photons in these massive leakers could escape from the ISM after considering photoelectric absorption, but they eventually get absorbed by the dust, resulting in much lower escape fractions. Fig.~\ref{fig:histogram} show a higher value of $\rm F_{900}/F_{1100}$ and also $\rm f_{esc, HI}$ in massive leakers on average than the LzLCS+ population. However, their $\rm f_{esc, tot}$ is significantly suppressed  and is almost equal to LzLCS+ detected sources, and that originates due a higher amount of dust attenuation.

To directly assess the relative impact of photoelectric absorption by neutral hydrogen and that of dust on the LyC escape fraction \citep[similar to ][]{xu23}, we calculate the escape fraction attributed solely to dust attenuation: $\rm f_{esc, dust}$ = $\rm \frac {f_{esc, tot} (Method \ 2)} { f_{esc, HI} (Method \ 3) }$. Fig.~\ref{fig:fesc-HI-dust} illustrates the difference between the escape fractions due to dust ($\rm f_{esc, dust}$) and neutral hydrogen ($\rm f_{esc, HI}$). Most massive leakers have $\rm f_{esc, HI} > f_{esc, dust}$, indicating that their LyC escape fractions are significantly reduced by dust attenuation.

\subsection{Comparison with other indirect \fesc \ diagnostics} 

The LzLCS+ survey and other studies in the past have shown that the LyC escape often exhibits significant correlation with different host galaxy properties, albeit with substantial scatter \citep{wang21, flury22b, izotov16, saldana-lopez22, chisholm22, jaskot24}. The observed correlations with \fesc, e.g. with $\rm SFR/Area$, $\rm f^{Ly\alpha}_{esc}$, Ly$\alpha$ EW, H$\beta$ EW, [OIII]/[OII] ($\rm O_{32}$), dust attenuation, [SII]-deficiency, $\beta$-slope and residual flux in Ly$\beta$  motivated some of these quantities to be used as potential indirect signposts for \fesc \ leakage. These indirect tracers are especially important for high redshift sources where directly measuring LyC flux density is impossible, for example, during the EoR \citep[e.g., ][]{jaskot24b}. 
In this analysis, we assess whether these correlations, previously established in traditional LyC-leaking galaxies from the LzLCS+ sample, also apply to the massive leaker population. For this comparison, we use three distinct proxies for LyC escape on the Y-axis.

\begin{figure*}
    \centering
    \includegraphics[width=\textwidth]{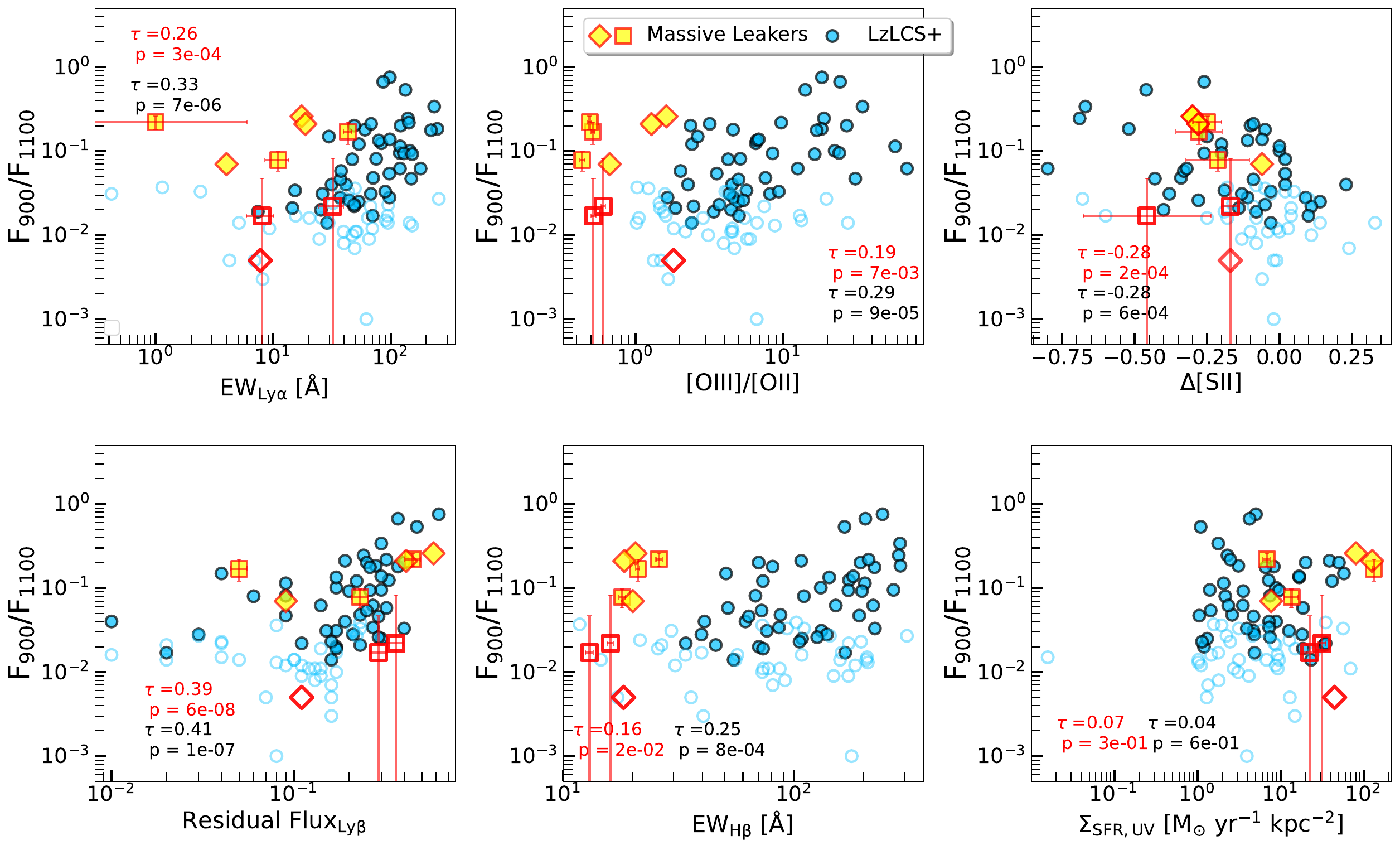}
    \caption{The correlations between \fesc \ measured from direct flux ratio: $\rm F_{900}/F_{1100}$ (Method 1, \S\ref{subsec:measure_lyc}) and various indirect tracers of \fesc\ are shown:  Ly$\alpha$ equivalent width (upper left), [OIII]/[OII] ratio (upper middle), $\Delta$ [SII] (upper right), Ly$\beta$ residual flux (lower left), H$\beta$ equivalent width (lower middle), and star formation rate surface density (lower right). Yellow diamonds represent the massive leakers from this study, while yellow squares denote those from \cite{wang19} and \cite{borthakur14}; blue circles show the LzLCS+ sources. Objects with upper limits are marked by hollow symbols. Each panel displays the Kendall’s $\tau$ coefficient and the null hypothesis probability (p) in the corner (red: combined LzLCS+ and massive leakers; black: LzLCS+ only). The correlation strength ($\tau$) generally decreases when including massive leakers (shown in red), compared to the LzLCS+ sample alone (black), except for $\Delta$[SII] and Ly$\beta$ residual flux, where the correlations remain preserved.  }
    \label{fig:indicators}
\end{figure*}

\begin{figure*}
    \centering
    \includegraphics[width=\textwidth]{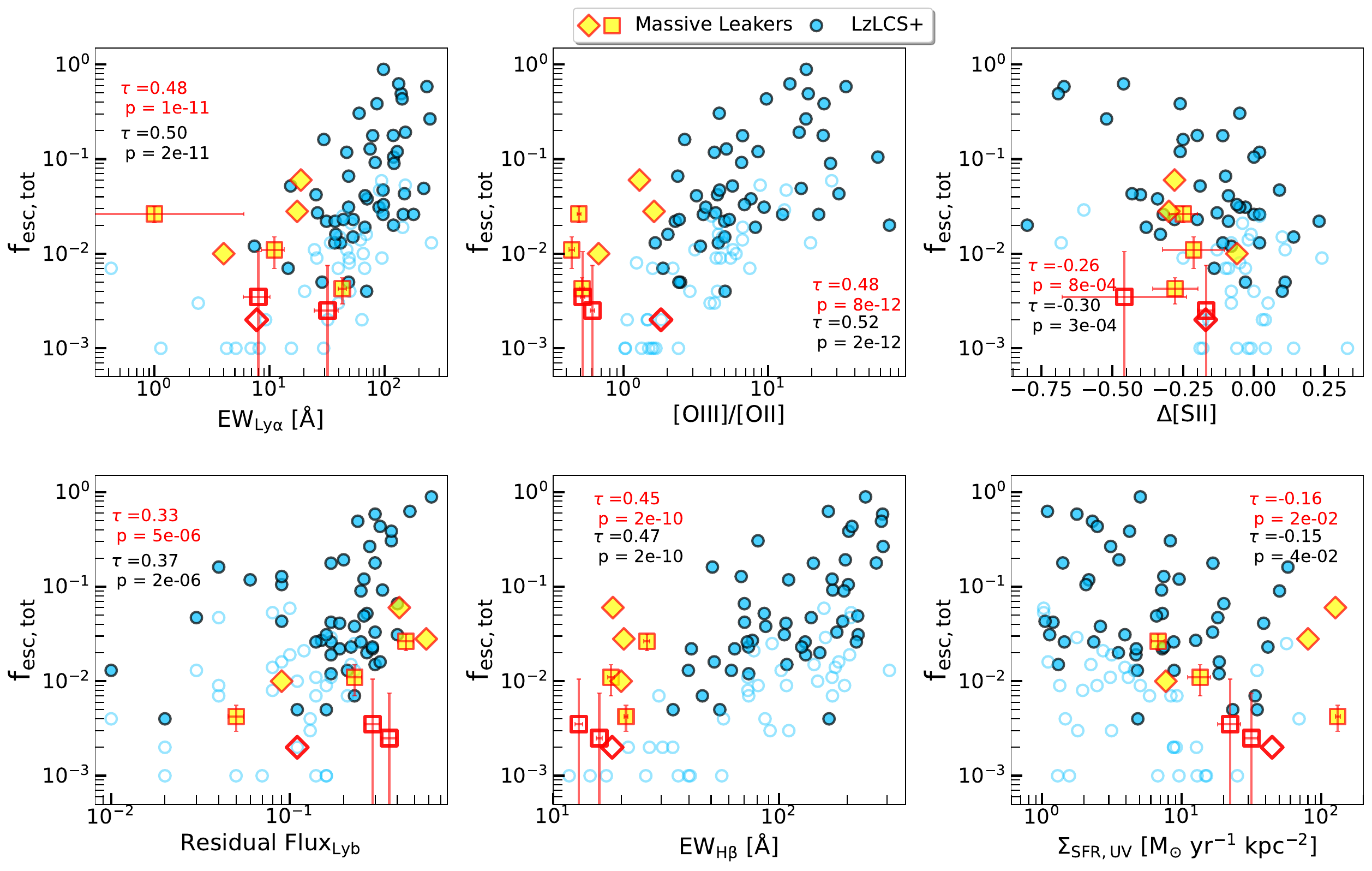}
    \caption{Correlations between the total \fesc \ computed from Method 2 (\S\ref{subsec:measure_lyc}) from UV SED fitting: $\rm f_{esc, tot}$ and various indirect tracers of \fesc:  Ly$\alpha$ equivalent width (upper left), [OIII]/[OII] ratio (upper middle), $\Delta$[SII] (upper right), Ly$\beta$ residual flux (lower left), H$\beta$ equivalent width (lower middle), and star formation rate surface density (lower right). Symbols are similar to the previous figure. 
    For each figure, the Kendall's $\tau$ coefficient between the x and y values and the probability of the null hypothesis (p) are shown at the corner (red: LzLCS+ \& massive leakers, black: only LzLCS+). In all the measured correlations shown above, the significance of the relation ($\tau$) diminishes when the massive leakers are taken into account (marked in red) compared to the correlation observed in the LzLCS+ sample (blue). The observed difference in correlations from Fig.~\ref{fig:indicators} represent is primarily due to the effect of dust.   }
    \label{fig:fesc_vs_indicators}
\end{figure*}

\begin{figure*}
    \centering
    \includegraphics[width=\textwidth]{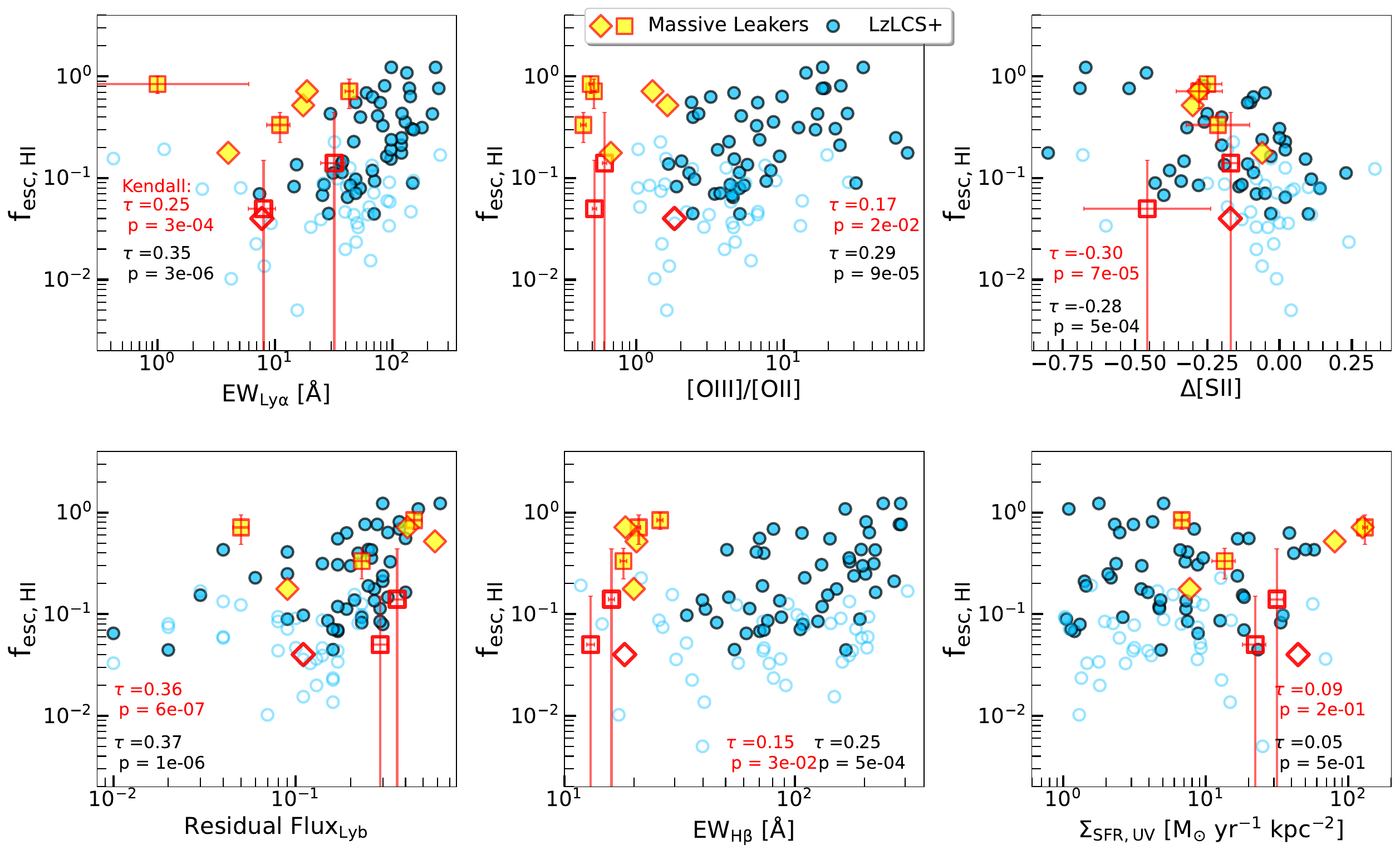}
    \caption{Correlations between the  \fesc \ due to HI absorption computed from Method 3 (\S\ref{subsec:measure_lyc}): $\rm f_{esc, HI}$ and various indirect tracers of \fesc:  Ly$\alpha$ equivalent width (upper left), [OIII]/[OII] ratio (upper middle), $\Delta$[SII] (upper right), Ly$\beta$ residual flux (lower left), H$\beta$ equivalent width (lower middle), and star formation rate surface density (lower right). Symbols are similar to the previous figures. 
    For each figure, the Kendall's $\tau$ coefficient between the x and y values and the probability of the null hypothesis (p) are shown at the corner (red: LzLCS+ \& massive leakers, black: only LzLCS+). In all the measured correlations shown above, the significance of the relation ($\tau$) diminishes when the massive leakers are taken into account (marked in red) compared to the correlation observed in the LzLCS+ sample (blue). The observed difference in correlations from Fig.~\ref{fig:indicators} represent is primarily due to the effect of dust.   }
    \label{fig:fescHI_vs_indicators}
\end{figure*}

Fig.~\ref{fig:indicators} displays the relationships between these indirect tracers and the LyC flux ratio $\rm F_{ \lambda LyC}/F_{\rm \lambda 1100}$ (or $\rm F_{900}/F_{1100}$), a model-independent, directly observed metric that is not directly affected by dust but does not directly measure $\rm f_{esc}$. 
Fig.~\ref{fig:fesc_vs_indicators} presents these tracers in relation to the more traditional LyC escape fraction $\rm f_{esc, tot}$, derived through stellar population modeling. This widely used measure of \fesc\ accounts for attenuation effect of LyC photons from both neutral hydrogen absorption and dust. 
Finally, Fig.~\ref{fig:fescHI_vs_indicators} shows the variation with $\rm f_{esc, HI}$, which isolates the effects of HI absorption alone. Each panel in these three figures displays the Kendall $\tau$ coefficient, indicating the strength of the correlation between the x and y values, along with the associated p-value representing the probability of the null hypothesis. Black-colored text represent the Kendall $\tau$ coefficients for the LzLCS+ sample alone (blue circles), while red-colored values include both the LzLCS+ sample and the massive leakers \citep[yellow diamonds: literature sample from ][yellow squares: this work]{wang19, borthakur14}. The difference in the $\tau$ values show whether the significance of the correlations decreases when massive leakers are included compared to the LzLCS+ sample alone. This, in turn, quantifies whether the massive leakers lie along the main relation shown by the LzLCS+, or whether they are significantly offset.

\begin{figure*}
    \centering
    \includegraphics[width=\textwidth]{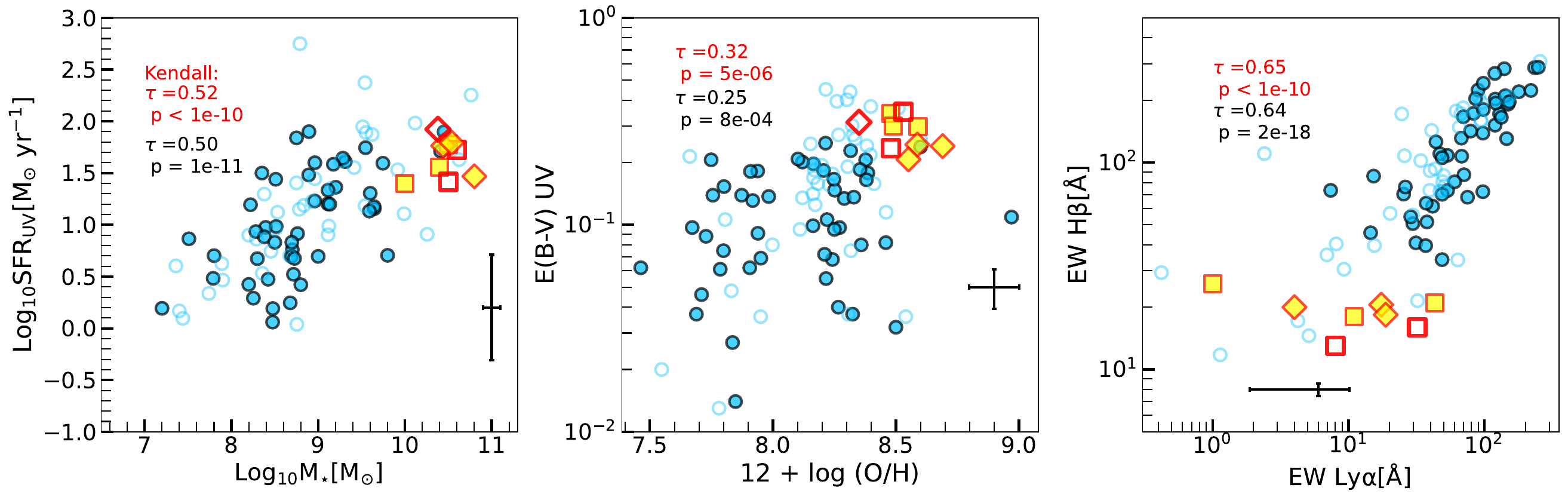}
    \caption{Comparison of host galaxy properties between the massive leaker population (yellow diamonds and squares) and the LzLCS+ sample (blue circles). The symbols represent the same meaning as the previous Figures.  The distributions are shown in three different parameter spaces: star formation rate derived from extinction-corrected UV 1500 Å luminosity ($\rm SFR_{UV}$) vs. stellar mass (left panel), UV extinction vs. metallicity (middle panel), and H$\beta$ equivalent width vs. Ly$\alpha$ equivalent width (right panel). Characteristic $1\sigma$ uncertainties of our sample are shown as black errorbar symbol. The massive leakers exhibit higher SFRs (but lower $\rm SFR/\rm{M_\star}$), greater dust extinction, higher metallicity, and weaker emission lines compared to typical LzLCS+ galaxies, indicating significant differences from traditional leakers in the literature.  }
    \label{fig:correlations}
\end{figure*}


The upper left panel in Fig~\ref{fig:indicators}  shows the relation between $\rm F_{900}/F_{1100}$ and the \lya \ equivalent width (EW) for the LzLCS+ sources (blue circles) and the massive leakers (yellow diamonds and squares). There is a strong positive correlation between the two quantities for the LzLCS+ sources. The Ly$\alpha$ line is produced from resonant scattering of \lya \ photons through the line of sight, and it is plausible for it to follow \fesc, since both \lya \ and LyC
emission depend on the distribution of neutral hydrogen in a
galaxy. Hence \lya \ EW is often used as an indirect \fesc \ estimate \citep[e.g., ][]{roy23}. 
However, the main result here is that the massive leaker sample is shifted towards lower \lya \ EW compared to the LzLCS+ galaxies. 
The mean \lya \ EW value for the former is 17 \AA, which is 4 times lower compared with the mean \lya \ EW = 70 \AA \ from the LzLCS+ sample. The lower middle panel is an analogous plot for the $\rm{H\beta}$ EW. Here the massive leakers are even more strongly offset from the LzLCS+ galaxies, with much weaker H$\beta$ emission.

There are several possible effects that may cause these offsets in \lya\ and $\rm{H\beta}$ EW. 
First, the massive leakers have significantly higher amounts of dust extinction that the typical LzLCS+ galaxies. This could significantly reduce the amount of scattered \lya\ photons that can escape. Since the amount of dust attenuation can be greater for the nebular emission-lines than for the starlight \citep{calzetti00, reddy16} this will reduce the EW of the $\rm{H\beta}$ line. Second, there may be a more pronounced contribution of starlight from non-ionizing stars in the massive leakers, particularly in the optical. This will further reduce the EW of $\rm{H\beta}$. Third, the $\rm F_{900}/F_{1100} $ values for our sources are so high that it is suppressing the amount of Balmer and Lyman photons that are intrinsically produced. Finally, the sources studied in these objects have higher stellar mass, and possibly includes a mix of old, evolved stellar population along with young stars, which can dilute the EW.

Fig~\ref{fig:indicators} (upper middle panel) illustrates another key result. The LzLCS+ sample shows a positive correlation where $\rm O_{32}$ increases with increasing $\rm F_{ \lambda LyC}$/F$_{\rm \lambda 1100}$. The simplest explanation for this positive trend between the ionization state of the ISM gas and the escape of the ionizing photons is that the column density of HI that can be photoionized is directly proportional to the ionization parameter (U - the ratio of the densities of ionizing photons and H), which is directly traced by $\rm O_{32}$. This makes it more likely for fully ionized channels to be created which allow LyC escape, consistent with the proposed ``density-bounded nebula'' scenario. 
Although the dependence of $\rm O_{32}$ on metallicity and ionization parameter may lead to a more complicated scenario 
\citep{sawant21, giammanco05, stasinska15}, nevertheless [OIII]/[OII] is widely used as an indirect tracer of \fesc. 
However, the massive leaker sample is in striking contrast to the LzLCS+ population in this regard, with a large offset towards low $\rm O_{32}$. The mean $\rm O_{32}$ is 0.5 which is roughly 16 times lower than the mean $\rm O_{32} $ observed from the the LzLCS+ sources. Thus, $\rm O_{32}$ doesn't emerge as the best indirect \fesc \  estimator when the massive leakers are included.
This indicates that the massive compact leakers might indeed be a separate population from the traditionally known LyC leaking galaxies with LyC escaping under entirely different ISM conditions.

The relative weakness of [SII] lines indicate a depleted (or lack of) partially ionized region in the outskirts of the Stromgren sphere of classical HII regions \citep{wang19}. These [SII]-weak regions trace channels that are optically thin to ionizing radiation, thus allowing the escape of LyC photons. Fig~\ref{fig:indicators} (upper right panel) shows the trend between $\Delta \rm [SII]$ and LyC emission extends to both the LzLCS+ sample and the massive leaker population, making it a good indirect indicator of LyC escape. However, consistent with previous studies  \citep{ramambason20, wang21}, the relationship shows substantial scatter, which could emerge from anisotropically escaping LyC photons, or line of sight variations, etc. Thus, even though [SII] deficiency can select LCE candidates, it is not obvious how \fesc \ can be directly derived from $\Delta$[SII].

Fig.~\ref{fig:indicators} (lower left panel) show a rather tight correlation between the residual flux (RF) in the core of the $\rm{Ly\beta}$ absorption-line and $\rm F_{ \lambda LyC}$/F$_{\rm \lambda 1100}$. Like $\Delta$[SII], this correlation is maintained in both type of leakers. The mean Ly$\beta$ RF for the massive leakers sample = 0.346, which is roughly two times higher than that of the LzLCS+ sample \citep{saldana-lopez22}. 
The residual fluxes (RF) at the bottom of the HI absorption lines are tracers of the covering fraction of the neutral hydrogen ($\rm C_f$).
In the case of a uniform HI screen geometry, the residual flux RF = 1 $-$ $\rm C_f$.
As illustrated by \cite{heckman11}, this observed trend is also consistent with picket-fence scenario where the covering factor of optically thick HI is not unity. Supernova-driven outflows or extremely intense LyC radiation (large U) can create channels or ``holes'' in the HI through which ionizing radiation can escape, leading to higher \fesc. 

Finally, we show in the lower right panel the relationship between $\rm F_{ \lambda LyC}$/F$_{\rm \lambda 1100}$. and the SFR per unit area (SFR/A). Massive leakers have a high value of star formation rate surface density (SFR/Area), with values ranging between $\sim \rm 10 - 100 \ M_{\odot} \ yr^{-1} \ kpc^{-2}$. This is typically an order of magnitude higher than the LzLCS+ leakers. The overall correlation show a significant scatter for both types of leakers.

In summary, the massive leakers align with the LzLCS+ trend and their measurements only in the parameter spaces of $\rm F_{900}/F_{1100}$ vs. [SII]-weakness and Ly$\beta$ residual flux. Thus the massive leakers lie along the main relation for these two indirect tracers, but show a distinct offset in all the remaining quantities. A similar pattern appears in Fig.~\ref{fig:fescHI_vs_indicators}, which compares these tracers with $\rm f_{esc, HI}$ instead of $\rm F_{900}/F_{1100}$. Both these \fesc\  quantities remove or minimize the effects of dust attenuation. Thus, the massive leaker population have continuity with the traditional leakers in the two parameters most related to the effects of LyC absorption by HI: [SII]-weakness and Ly$\beta$ residual flux. Fig.~\ref{fig:fesc_vs_indicators}, on the other hand, replaces  $\rm F_{900}/F_{1100}$ with $\rm f_{esc, HI}$ in the vertical axis, which is sensitive to both dust and HI. This comparison reveals that all quantities for the massive leakers distinctly deviate from the LzLCS+ sources, indicating that dust attenuation shifts these massive leakers away from the earlier trends observed in traditional leakers.



\subsection{Are these a different class of leakers?} \label{subsec:results_dcos}


We have discussed in the previous section that the massive leakers show some notable differences from the other LyC leaking galaxies in regard to indirect \fesc \ indicators. Particularly, they do not follow the trends of leakiness correlating with primarily in $\rm O_{32}$, or the \lya \ and $\rm{H\beta}$ EW.  Here we evaluate the relationship between a few other host galaxy properties for the massive leakers, and compare them to the LzLCS+ galaxies. 
Fig.~\ref{fig:correlations} shows the galaxy distributions in three different planes -- star formation rate derived from the extinction-corrected UV 1500 \AA \ luminosity ($\rm SFR_{UV}$) vs. stellar mass (left panel), UV extinction vs. metallicity (middle panel), and  H$\beta$ EW vs \lya \ EW (right panel). 
It is clear that 
our sources are more massive, have higher SFR (but lower $\rm SFR/\rm{M_\star}$), and also very high SFR/Area than the typical LzLCS+ galaxies. These also have significantly higher dust extinction and higher metallicites than the LzLCS+ galaxies, but comparatively weaker emission lines. Thus, these massive leakers differ significantly from the population of `traditional' leakers in the literature. This is further supported by the different emission-line properties of the ISM described in the previous section.

We agree with the hypothesis put forward by \cite{heckman11, borthakur14, alexandroff15, wang19, flury22b, jaskot24} that the LyC escape in these massive leakers is made possible by the extreme feedback effects produced by an extremely massive and compact starburst \citep{flury24}. In the three massive leakers discussed by \cite{borthakur14} and \cite{wang19}, the UV light from starburst was dominated by highly compact (marginally resolved by
HST),  massive objects located at or near the
galactic nucleus. These were dubbed Dominant Central Objects \citep[DCOs;][]{overzier09}. In this study, we have confirmed LyC emission in three out of the five sources. All three have very high SFR/Area, but only J0121 has the morphology of a compact DCO  as shown in Fig.~\ref{fig:image}. The rest have comparatively extended morphology.  This suggest that the more general property of very high SFR/Area is responsible for enabling the escape of LyC photons. The typically large values of the RF in the $\rm{Ly\beta}$ absorption-line in the massive leakers is fully consistent with a ``picket fence'' geometry for the HI with holes that are fully ionized, enabling LyC escape and also reducing the relative amount of partially-ionized gas traced by [SII]. It is noteworthy that high SFR/A is correlated with high velocities in the starburst-generated galactic winds \citep[e.g., ][]{heckman15, heckman17, davis23}. This suggests that the ionized holes may be created - at least in part - as a result of the extremely high ram pressure associated with the wind fluid. Indeed, \cite{amorin24} recently found correlations between the velocity of these feedback driven winds and the LyC escape fraction, while \cite{flury24} explored the in-depth connections between the different types of feedback mechanism and the LyC escape.  Our low-resolution HST/COS data could not confirm blueshifts in the absorption lines indicative of outflows. We are currently planning follow-up observations with the Large Binocular Telescope to obtain higher spectral resolution data. This analysis will be addressed in future work.



\section{Conclusions} \label{sec:conclusion}
We present HST/COS spectroscopy observations of five low-redshift (z $\sim$ 0.3) starburst galaxies, selected for their high stellar mass ($\rm M_{\odot} > 10^{10}$) and relatively weak [S II] 6717\AA,6731\AA\ nebular emission lines compared to typical star-forming galaxies. The weakness in [S II] is hypothesized to indicate galaxies that are optically thin to ionizing radiation. Recently, several high-redshift galaxies (z $>$ 5) observed with JWST have also shown extremely weak [S II] emission \citep{sanders23, cameron23}, reinforcing the potential of weak [S II] as a marker of LyC escape in high-z galaxies.

Significant LyC flux was detected in three of the five galaxies, with dust-corrected escape fractions ($\rm f_{esc, HI}$) of 71\%, 84\%, and 33\%, indicating the proportion of LyC photons that would escape in dust-free conditions. However, dust substantially reduces the escaping LyC photon count. Consequently, the total escape fraction ($\rm f_{esc, tot}$)—the fraction of photons that actually escape and reach the IGM after surviving both neutral hydrogen absorption and dust extinction—is considerably lower, ranging from 1-3\%.

We compared these  [S II]-weak galaxies to other known classes of LyC-leaking galaxies. In contrast to the low-redshift traditional leakers from the LzLCS+ sample, these galaxies exhibit higher stellar masses, greater metallicities, more dust extinction, lower ionization states (as indicated by [OIII]/[OII]), smaller Ly$\alpha$ and H$\beta$ equivalent widths, and higher star formation rate surface densities. These results indicate that the massive leakers in our sample constitute a distinct population from the traditional LzLCS+ leakers, suggesting the presence of multiple mechanisms facilitating LyC photon escape. The characteristics of these massive leakers align with a ``picket-fence'' model, in which intense feedback creates channels or holes in the neutral hydrogen, allowing LyC photons to escape through these openings. With JWST and other recent ground-based studies uncovering an increasing number of massive ($>\rm 10^{10} \ M_{\odot}$) galaxies at high redshifts ($z \geq 3$) \citep[][]{labbe23, long24, carnall23, nanayakkara24, carnall20, schreiber18, glazebrook17}, this population of massive leaky starbursts will expand the range of galaxy properties available for LyC-leakage studies. This also enhances our ability to use low-redshift leakers as local analogs for investigating the physical mechanisms that facilitate LyC photon escape at higher redshifts.  It points to the likelihood that various physical conditions and mechanisms contribute to LyC leakage.

\software{Astropy \citep{astropy18, astropy22},  
          Scipy \citep{scipy20}, 
          HST STScI pipeline CALCOS \href{https://github.com/spacetelescope/calcos}{https://github.com/spacetelescope/calcos},
          FAINTCOS \citep{worseck16, makan21}, 
          FICUS \citep{saldana-lopez22, chisholm19}
          }.

\begin{acknowledgements}
M.J.H. is supported by the Swedish Research Council
(Vetenskapsr\aa{}det) and is Fellow of the Knut \& Alice Wallenberg
Foundation.
\end{acknowledgements}

\bibliography{main}

\begin{thebibliography}{}
\expandafter\ifx\csname natexlab\endcsname\relax\def\natexlab#1{#1}\fi
\providecommand{\url}[1]{\href{#1}{#1}}
\providecommand{\dodoi}[1]{doi:~\href{http://doi.org/#1}{\nolinkurl{#1}}}
\providecommand{\doeprint}[1]{\href{http://ascl.net/#1}{\nolinkurl{http://ascl.net/#1}}}
\providecommand{\doarXiv}[1]{\href{https://arxiv.org/abs/#1}{\nolinkurl{https://arxiv.org/abs/#1}}}

\bibitem[{{Alavi} {et~al.}(2020){Alavi}, {Colbert}, {Teplitz}, {Siana}, {Scarlata}, {Rutkowski}, {Mehta}, {Henry}, {Dai}, {Haardt}, \& {Bagley}}]{alavi20}
{Alavi}, A., {Colbert}, J., {Teplitz}, H.~I., {et~al.} 2020, \apj, 904, 59, \dodoi{10.3847/1538-4357/abbd43}

\bibitem[{{Alexandroff} {et~al.}(2015){Alexandroff}, {Heckman}, {Borthakur}, {Overzier}, \& {Leitherer}}]{alexandroff15}
{Alexandroff}, R.~M., {Heckman}, T.~M., {Borthakur}, S., {Overzier}, R., \& {Leitherer}, C. 2015, \apj, 810, 104, \dodoi{10.1088/0004-637X/810/2/104}

\bibitem[{{Amor{\'\i}n} {et~al.}(2024){Amor{\'\i}n}, {Rodr{\'\i}guez-Henr{\'\i}quez}, {Fern{\'a}ndez}, {V{\'\i}lchez}, {Marques-Chaves}, {Schaerer}, {Izotov}, {Firpo}, {Guseva}, {Jaskot}, {Komarova}, {Mu{\~n}oz-Vergara}, {Oey}, {Bait}, {Carr}, {Chisholm}, {Ferguson}, {Flury}, {Giavalisco}, {Hayes}, {Henry}, {Ji}, {King}, {Leclercq}, {{\"O}stlin}, {Pentericci}, {Saldana-Lopez}, {Thuan}, {Trebitsch}, {Wang}, {Worseck}, \& {Xu}}]{amorin24}
{Amor{\'\i}n}, R.~O., {Rodr{\'\i}guez-Henr{\'\i}quez}, M., {Fern{\'a}ndez}, V., {et~al.} 2024, \aap, 682, L25, \dodoi{10.1051/0004-6361/202449175}

\bibitem[{{Astropy Collaboration} {et~al.}(2018){Astropy Collaboration}, {Price-Whelan}, {Sip{\H{o}}cz}, {G{\"u}nther}, {Lim}, {Crawford}, {Conseil}, {Shupe}, {Craig}, {Dencheva}, {Ginsburg}, {VanderPlas}, {Bradley}, {P{\'e}rez-Su{\'a}rez}, {de Val-Borro}, {Aldcroft}, {Cruz}, {Robitaille}, {Tollerud}, {Ardelean}, {Babej}, {Bach}, {Bachetti}, {Bakanov}, {Bamford}, {Barentsen}, {Barmby}, {Baumbach}, {Berry}, {Biscani}, {Boquien}, {Bostroem}, {Bouma}, {Brammer}, {Bray}, {Breytenbach}, {Buddelmeijer}, {Burke}, {Calderone}, {Cano Rodr{\'\i}guez}, {Cara}, {Cardoso}, {Cheedella}, {Copin}, {Corrales}, {Crichton}, {D'Avella}, {Deil}, {Depagne}, {Dietrich}, {Donath}, {Droettboom}, {Earl}, {Erben}, {Fabbro}, {Ferreira}, {Finethy}, {Fox}, {Garrison}, {Gibbons}, {Goldstein}, {Gommers}, {Greco}, {Greenfield}, {Groener}, {Grollier}, {Hagen}, {Hirst}, {Homeier}, {Horton}, {Hosseinzadeh}, {Hu}, {Hunkeler}, {Ivezi{\'c}}, {Jain}, {Jenness}, {Kanarek}, {Kendrew}, {Kern}, {Kerzendorf}, {Khvalko}, {King}, {Kirkby}, {Kulkarni},
  {Kumar}, {Lee}, {Lenz}, {Littlefair}, {Ma}, {Macleod}, {Mastropietro}, {McCully}, {Montagnac}, {Morris}, {Mueller}, {Mumford}, {Muna}, {Murphy}, {Nelson}, {Nguyen}, {Ninan}, {N{\"o}the}, {Ogaz}, {Oh}, {Parejko}, {Parley}, {Pascual}, {Patil}, {Patil}, {Plunkett}, {Prochaska}, {Rastogi}, {Reddy Janga}, {Sabater}, {Sakurikar}, {Seifert}, {Sherbert}, {Sherwood-Taylor}, {Shih}, {Sick}, {Silbiger}, {Singanamalla}, {Singer}, {Sladen}, {Sooley}, {Sornarajah}, {Streicher}, {Teuben}, {Thomas}, {Tremblay}, {Turner}, {Terr{\'o}n}, {van Kerkwijk}, {de la Vega}, {Watkins}, {Weaver}, {Whitmore}, {Woillez}, {Zabalza}, \& {Astropy Contributors}}]{astropy18}
{Astropy Collaboration}, {Price-Whelan}, A.~M., {Sip{\H{o}}cz}, B.~M., {et~al.} 2018, \aj, 156, 123, \dodoi{10.3847/1538-3881/aabc4f}

\bibitem[{{Astropy Collaboration} {et~al.}(2022){Astropy Collaboration}, {Price-Whelan}, {Lim}, {Earl}, {Starkman}, {Bradley}, {Shupe}, {Patil}, {Corrales}, {Brasseur}, {N{\"o}the}, {Donath}, {Tollerud}, {Morris}, {Ginsburg}, {Vaher}, {Weaver}, {Tocknell}, {Jamieson}, {van Kerkwijk}, {Robitaille}, {Merry}, {Bachetti}, {G{\"u}nther}, {Aldcroft}, {Alvarado-Montes}, {Archibald}, {B{\'o}di}, {Bapat}, {Barentsen}, {Baz{\'a}n}, {Biswas}, {Boquien}, {Burke}, {Cara}, {Cara}, {Conroy}, {Conseil}, {Craig}, {Cross}, {Cruz}, {D'Eugenio}, {Dencheva}, {Devillepoix}, {Dietrich}, {Eigenbrot}, {Erben}, {Ferreira}, {Foreman-Mackey}, {Fox}, {Freij}, {Garg}, {Geda}, {Glattly}, {Gondhalekar}, {Gordon}, {Grant}, {Greenfield}, {Groener}, {Guest}, {Gurovich}, {Handberg}, {Hart}, {Hatfield-Dodds}, {Homeier}, {Hosseinzadeh}, {Jenness}, {Jones}, {Joseph}, {Kalmbach}, {Karamehmetoglu}, {Ka{\l}uszy{\'n}ski}, {Kelley}, {Kern}, {Kerzendorf}, {Koch}, {Kulumani}, {Lee}, {Ly}, {Ma}, {MacBride}, {Maljaars}, {Muna}, {Murphy}, {Norman},
  {O'Steen}, {Oman}, {Pacifici}, {Pascual}, {Pascual-Granado}, {Patil}, {Perren}, {Pickering}, {Rastogi}, {Roulston}, {Ryan}, {Rykoff}, {Sabater}, {Sakurikar}, {Salgado}, {Sanghi}, {Saunders}, {Savchenko}, {Schwardt}, {Seifert-Eckert}, {Shih}, {Jain}, {Shukla}, {Sick}, {Simpson}, {Singanamalla}, {Singer}, {Singhal}, {Sinha}, {Sip{\H{o}}cz}, {Spitler}, {Stansby}, {Streicher}, {{\v{S}}umak}, {Swinbank}, {Taranu}, {Tewary}, {Tremblay}, {de Val-Borro}, {Van Kooten}, {Vasovi{\'c}}, {Verma}, {de Miranda Cardoso}, {Williams}, {Wilson}, {Winkel}, {Wood-Vasey}, {Xue}, {Yoachim}, {Zhang}, {Zonca}, \& {Astropy Project Contributors}}]{astropy22}
{Astropy Collaboration}, {Price-Whelan}, A.~M., {Lim}, P.~L., {et~al.} 2022, \apj, 935, 167, \dodoi{10.3847/1538-4357/ac7c74}

\bibitem[{{Atek} {et~al.}(2024){Atek}, {Labb{\'e}}, {Furtak}, {Chemerynska}, {Fujimoto}, {Setton}, {Miller}, {Oesch}, {Bezanson}, {Price}, {Dayal}, {Zitrin}, {Kokorev}, {Weaver}, {Brammer}, {Dokkum}, {Williams}, {Cutler}, {Feldmann}, {Fudamoto}, {Greene}, {Leja}, {Maseda}, {Muzzin}, {Pan}, {Papovich}, {Nelson}, {Nanayakkara}, {Stark}, {Stefanon}, {Suess}, {Wang}, \& {Whitaker}}]{atek24}
{Atek}, H., {Labb{\'e}}, I., {Furtak}, L.~J., {et~al.} 2024, \nat, 626, 975, \dodoi{10.1038/s41586-024-07043-6}

\bibitem[{{Ba{\~n}ados} {et~al.}(2018){Ba{\~n}ados}, {Carilli}, {Walter}, {Momjian}, {Decarli}, {Farina}, {Mazzucchelli}, \& {Venemans}}]{banados18}
{Ba{\~n}ados}, E., {Carilli}, C., {Walter}, F., {et~al.} 2018, \apjl, 861, L14, \dodoi{10.3847/2041-8213/aac511}

\bibitem[{{Baldwin} {et~al.}(1981){Baldwin}, {Phillips}, \& {Terlevich}}]{baldwin81}
{Baldwin}, J.~A., {Phillips}, M.~M., \& {Terlevich}, R. 1981, \pasp, 93, 5, \dodoi{10.1086/130766}

\bibitem[{{Becker} {et~al.}(2001){Becker}, {Fan}, {White}, {Strauss}, {Narayanan}, {Lupton}, {Gunn}, {Annis}, {Bahcall}, {Brinkmann}, {Connolly}, {Csabai}, {Czarapata}, {Doi}, {Heckman}, {Hennessy}, {Ivezi{\'c}}, {Knapp}, {Lamb}, {McKay}, {Munn}, {Nash}, {Nichol}, {Pier}, {Richards}, {Schneider}, {Stoughton}, {Szalay}, {Thakar}, \& {York}}]{becker01}
{Becker}, R.~H., {Fan}, X., {White}, R.~L., {et~al.} 2001, \aj, 122, 2850, \dodoi{10.1086/324231}

\bibitem[{{Bergvall} {et~al.}(2006){Bergvall}, {Zackrisson}, {Andersson}, {Arnberg}, {Masegosa}, \& {{\"O}stlin}}]{bergvall06}
{Bergvall}, N., {Zackrisson}, E., {Andersson}, B.~G., {et~al.} 2006, \aap, 448, 513, \dodoi{10.1051/0004-6361:20053788}

\bibitem[{{Borthakur} {et~al.}(2014){Borthakur}, {Heckman}, {Leitherer}, \& {Overzier}}]{borthakur14}
{Borthakur}, S., {Heckman}, T.~M., {Leitherer}, C., \& {Overzier}, R.~A. 2014, Science, 346, 216, \dodoi{10.1126/science.1254214}

\bibitem[{{Bouwens} {et~al.}(2016){Bouwens}, {Smit}, {Labb{\'e}}, {Franx}, {Caruana}, {Oesch}, {Stefanon}, \& {Rasappu}}]{bouwens16}
{Bouwens}, R.~J., {Smit}, R., {Labb{\'e}}, I., {et~al.} 2016, \apj, 831, 176, \dodoi{10.3847/0004-637X/831/2/176}

\bibitem[{{Bouwens} {et~al.}(2011){Bouwens}, {Illingworth}, {Oesch}, {Labb{\'e}}, {Trenti}, {van Dokkum}, {Franx}, {Stiavelli}, {Carollo}, {Magee}, \& {Gonzalez}}]{bouwens11}
{Bouwens}, R.~J., {Illingworth}, G.~D., {Oesch}, P.~A., {et~al.} 2011, \apj, 737, 90, \dodoi{10.1088/0004-637X/737/2/90}

\bibitem[{{Calzetti} {et~al.}(2000){Calzetti}, {Armus}, {Bohlin}, {Kinney}, {Koornneef}, \& {Storchi-Bergmann}}]{calzetti00}
{Calzetti}, D., {Armus}, L., {Bohlin}, R.~C., {et~al.} 2000, \apj, 533, 682, \dodoi{10.1086/308692}

\bibitem[{{Cameron} {et~al.}(2023){Cameron}, {Saxena}, {Bunker}, {D'Eugenio}, {Carniani}, {Maiolino}, {Curtis-Lake}, {Ferruit}, {Jakobsen}, {Arribas}, {Bonaventura}, {Charlot}, {Chevallard}, {Curti}, {Looser}, {Maseda}, {Rawle}, {Rodr{\'\i}guez Del Pino}, {Smit}, {{\"U}bler}, {Willott}, {Witstok}, {Egami}, {Eisenstein}, {Johnson}, {Hainline}, {Rieke}, {Robertson}, {Stark}, {Tacchella}, {Williams}, {Bhatawdekar}, {Bowler}, {Boyett}, {Circosta}, {Helton}, {Jones}, {Kumari}, {Ji}, {Nelson}, {Parlanti}, {Sandles}, {Scholtz}, \& {Sun}}]{cameron23}
{Cameron}, A.~J., {Saxena}, A., {Bunker}, A.~J., {et~al.} 2023, arXiv e-prints, arXiv:2302.04298, \dodoi{10.48550/arXiv.2302.04298}

\bibitem[{{Carnall} {et~al.}(2020){Carnall}, {Walker}, {McLure}, {Dunlop}, {McLeod}, {Cullen}, {Wild}, {Amorin}, {Bolzonella}, {Castellano}, {Cimatti}, {Cucciati}, {Fontana}, {Gargiulo}, {Garilli}, {Jarvis}, {Pentericci}, {Pozzetti}, {Zamorani}, {Calabro}, {Hathi}, \& {Koekemoer}}]{carnall20}
{Carnall}, A.~C., {Walker}, S., {McLure}, R.~J., {et~al.} 2020, \mnras, 496, 695, \dodoi{10.1093/mnras/staa1535}

\bibitem[{{Carnall} {et~al.}(2023){Carnall}, {McLeod}, {McLure}, {Dunlop}, {Begley}, {Cullen}, {Donnan}, {Hamadouche}, {Jewell}, {Jones}, {Pollock}, \& {Wild}}]{carnall23}
{Carnall}, A.~C., {McLeod}, D.~J., {McLure}, R.~J., {et~al.} 2023, \mnras, 520, 3974, \dodoi{10.1093/mnras/stad369}

\bibitem[{{Chisholm} {et~al.}(2019){Chisholm}, {Rigby}, {Bayliss}, {Berg}, {Dahle}, {Gladders}, \& {Sharon}}]{chisholm19}
{Chisholm}, J., {Rigby}, J.~R., {Bayliss}, M., {et~al.} 2019, \apj, 882, 182, \dodoi{10.3847/1538-4357/ab3104}

\bibitem[{{Chisholm} {et~al.}(2022){Chisholm}, {Saldana-Lopez}, {Flury}, {Schaerer}, {Jaskot}, {Amor{\'\i}n}, {Atek}, {Finkelstein}, {Fleming}, {Ferguson}, {Fern{\'a}ndez}, {Giavalisco}, {Hayes}, {Heckman}, {Henry}, {Ji}, {Marques-Chaves}, {Mauerhofer}, {McCandliss}, {Oey}, {{\"O}stlin}, {Rutkowski}, {Scarlata}, {Thuan}, {Trebitsch}, {Wang}, {Worseck}, \& {Xu}}]{chisholm22}
{Chisholm}, J., {Saldana-Lopez}, A., {Flury}, S., {et~al.} 2022, \mnras, 517, 5104, \dodoi{10.1093/mnras/stac2874}

\bibitem[{{Davis} {et~al.}(2023){Davis}, {Tremonti}, {Swiggum}, {Moustakas}, {Diamond-Stanic}, {Coil}, {Geach}, {Hickox}, {Perrotta}, {Petter}, {Rudnick}, {Rupke}, {Sell}, \& {Whalen}}]{davis23}
{Davis}, J.~D., {Tremonti}, C.~A., {Swiggum}, C.~N., {et~al.} 2023, \apj, 951, 105, \dodoi{10.3847/1538-4357/accbbf}

\bibitem[{{Deharveng} {et~al.}(2001){Deharveng}, {Buat}, {Le Brun}, {Milliard}, {Kunth}, {Shull}, \& {Gry}}]{deharveng01}
{Deharveng}, J.~M., {Buat}, V., {Le Brun}, V., {et~al.} 2001, \aap, 375, 805, \dodoi{10.1051/0004-6361:20010920}

\bibitem[{{Fan} {et~al.}(2006){Fan}, {Carilli}, \& {Keating}}]{fan06}
{Fan}, X., {Carilli}, C.~L., \& {Keating}, B. 2006, \araa, 44, 415, \dodoi{10.1146/annurev.astro.44.051905.092514}

\bibitem[{{Feldman} \& {Cousins}(1998)}]{feldman98}
{Feldman}, G.~J., \& {Cousins}, R.~D. 1998, \prd, 57, 3873, \dodoi{10.1103/PhysRevD.57.3873}

\bibitem[{{Ferland} {et~al.}(2017){Ferland}, {Chatzikos}, {Guzm{\'a}n}, {Lykins}, {van Hoof}, {Williams}, {Abel}, {Badnell}, {Keenan}, {Porter}, \& {Stancil}}]{ferland17}
{Ferland}, G.~J., {Chatzikos}, M., {Guzm{\'a}n}, F., {et~al.} 2017, \rmxaa, 53, 385, \dodoi{10.48550/arXiv.1705.10877}

\bibitem[{{Finkelstein} {et~al.}(2019){Finkelstein}, {D'Aloisio}, {Paardekooper}, {Ryan}, {Behroozi}, {Finlator}, {Livermore}, {Upton Sanderbeck}, {Dalla Vecchia}, \& {Khochfar}}]{finkelstein19}
{Finkelstein}, S.~L., {D'Aloisio}, A., {Paardekooper}, J.-P., {et~al.} 2019, \apj, 879, 36, \dodoi{10.3847/1538-4357/ab1ea8}

\bibitem[{{Fletcher} {et~al.}(2019){Fletcher}, {Tang}, {Robertson}, {Nakajima}, {Ellis}, {Stark}, \& {Inoue}}]{fletcher19}
{Fletcher}, T.~J., {Tang}, M., {Robertson}, B.~E., {et~al.} 2019, \apj, 878, 87, \dodoi{10.3847/1538-4357/ab2045}

\bibitem[{Flury(2025)}]{feldcous}
Flury, S. 2025, sflury/FeldCous: FeldCous v1.0.0, v1.0.0,  Zenodo, \dodoi{10.5281/zenodo.14857987}

\bibitem[{{Flury} {et~al.}(2022{\natexlab{a}}){Flury}, {Jaskot}, {Ferguson}, {Worseck}, {Makan}, {Chisholm}, {Saldana-Lopez}, {Schaerer}, {McCandliss}, {Wang}, {Ford}, {Heckman}, {Ji}, {Giavalisco}, {Amorin}, {Atek}, {Blaizot}, {Borthakur}, {Carr}, {Castellano}, {Cristiani}, {De Barros}, {Dickinson}, {Finkelstein}, {Fleming}, {Fontanot}, {Garel}, {Grazian}, {Hayes}, {Henry}, {Mauerhofer}, {Micheva}, {Oey}, {Ostlin}, {Papovich}, {Pentericci}, {Ravindranath}, {Rosdahl}, {Rutkowski}, {Santini}, {Scarlata}, {Teplitz}, {Thuan}, {Trebitsch}, {Vanzella}, {Verhamme}, \& {Xu}}]{flury22}
{Flury}, S.~R., {Jaskot}, A.~E., {Ferguson}, H.~C., {et~al.} 2022{\natexlab{a}}, \apjs, 260, 1, \dodoi{10.3847/1538-4365/ac5331}

\bibitem[{{Flury} {et~al.}(2022{\natexlab{b}}){Flury}, {Jaskot}, {Ferguson}, {Worseck}, {Makan}, {Chisholm}, {Saldana-Lopez}, {Schaerer}, {McCandliss}, {Xu}, {Wang}, {Oey}, {Ford}, {Heckman}, {Ji}, {Giavalisco}, {Amor{\'\i}n}, {Atek}, {Blaizot}, {Borthakur}, {Carr}, {Castellano}, {De Barros}, {Dickinson}, {Finkelstein}, {Fleming}, {Fontanot}, {Garel}, {Grazian}, {Hayes}, {Henry}, {Mauerhofer}, {Micheva}, {Ostlin}, {Papovich}, {Pentericci}, {Ravindranath}, {Rosdahl}, {Rutkowski}, {Santini}, {Scarlata}, {Teplitz}, {Thuan}, {Trebitsch}, {Vanzella}, \& {Verhamme}}]{flury22b}
---. 2022{\natexlab{b}}, \apj, 930, 126, \dodoi{10.3847/1538-4357/ac61e4}

\bibitem[{{Flury} {et~al.}(2024){Flury}, {Jaskot}, {Saldana-Lopez}, {Oey}, {Chisholm}, {Amor{\'\i}n}, {Bait}, {Borthakur}, {Carr}, {Ferguson}, {Giavalisco}, {Hayes}, {Heckman}, {Henry}, {Ji}, {Komarova}, {Leclercq}, {Le Reste}, {McCandliss}, {Marques-Chaves}, {{\"O}stlin}, {Pentericci}, {Ravindranath}, {Rutkowski}, {Scarlata}, {Schaerer}, {Thuan}, {Trebitsch}, {Vanzella}, {Verhamme}, {Wang}, {Worseck}, \& {Xu}}]{flury24}
{Flury}, S.~R., {Jaskot}, A.~E., {Saldana-Lopez}, A., {et~al.} 2024, arXiv e-prints, arXiv:2409.12118, \dodoi{10.48550/arXiv.2409.12118}

\bibitem[{{Gazagnes} {et~al.}(2020){Gazagnes}, {Chisholm}, {Schaerer}, {Verhamme}, \& {Izotov}}]{gazagnes20}
{Gazagnes}, S., {Chisholm}, J., {Schaerer}, D., {Verhamme}, A., \& {Izotov}, Y. 2020, \aap, 639, A85, \dodoi{10.1051/0004-6361/202038096}

\bibitem[{{Giammanco} {et~al.}(2005){Giammanco}, {Beckman}, \& {Cedr{\'e}s}}]{giammanco05}
{Giammanco}, C., {Beckman}, J.~E., \& {Cedr{\'e}s}, B. 2005, \aap, 438, 599, \dodoi{10.1051/0004-6361:20042268}

\bibitem[{{Glazebrook} {et~al.}(2017){Glazebrook}, {Schreiber}, {Labb{\'e}}, {Nanayakkara}, {Kacprzak}, {Oesch}, {Papovich}, {Spitler}, {Straatman}, {Tran}, \& {Yuan}}]{glazebrook17}
{Glazebrook}, K., {Schreiber}, C., {Labb{\'e}}, I., {et~al.} 2017, \nat, 544, 71, \dodoi{10.1038/nature21680}

\bibitem[{{Green} {et~al.}(2012){Green}, {Froning}, {Osterman}, {Ebbets}, {Heap}, {Leitherer}, {Linsky}, {Savage}, {Sembach}, {Shull}, {Siegmund}, {Snow}, {Spencer}, {Stern}, {Stocke}, {Welsh}, {B{\'e}land}, {Burgh}, {Danforth}, {France}, {Keeney}, {McPhate}, {Penton}, {Andrews}, {Brownsberger}, {Morse}, \& {Wilkinson}}]{green12}
{Green}, J.~C., {Froning}, C.~S., {Osterman}, S., {et~al.} 2012, \apj, 744, 60, \dodoi{10.1088/0004-637X/744/1/6010.1086/141956}

\bibitem[{{Hayes} {et~al.}(2025){Hayes}, {Saldana-Lopez}, {Citro}, {James}, {Mingozzi}, {Scarlata}, {Martinez}, \& {Berg}}]{hayes25}
{Hayes}, M.~J., {Saldana-Lopez}, A., {Citro}, A., {et~al.} 2025, \apj, 982, 14, \dodoi{10.3847/1538-4357/adaea1}

\bibitem[{{Heckman} {et~al.}(2017){Heckman}, {Borthakur}, {Wild}, {Schiminovich}, \& {Bordoloi}}]{heckman17}
{Heckman}, T., {Borthakur}, S., {Wild}, V., {Schiminovich}, D., \& {Bordoloi}, R. 2017, \apj, 846, 151, \dodoi{10.3847/1538-4357/aa80dc}

\bibitem[{{Heckman} {et~al.}(2015){Heckman}, {Alexandroff}, {Borthakur}, {Overzier}, \& {Leitherer}}]{heckman15}
{Heckman}, T.~M., {Alexandroff}, R.~M., {Borthakur}, S., {Overzier}, R., \& {Leitherer}, C. 2015, \apj, 809, 147, \dodoi{10.1088/0004-637X/809/2/147}

\bibitem[{{Heckman} {et~al.}(2005){Heckman}, {Hoopes}, {Seibert}, {Martin}, {Salim}, {Rich}, {Kauffmann}, {Charlot}, {Barlow}, {Bianchi}, {Byun}, {Donas}, {Forster}, {Friedman}, {Jelinsky}, {Lee}, {Madore}, {Malina}, {Milliard}, {Morrissey}, {Neff}, {Schiminovich}, {Siegmund}, {Small}, {Szalay}, {Welsh}, \& {Wyder}}]{heckman05}
{Heckman}, T.~M., {Hoopes}, C.~G., {Seibert}, M., {et~al.} 2005, \apjl, 619, L35, \dodoi{10.1086/425979}

\bibitem[{{Heckman} {et~al.}(2011){Heckman}, {Borthakur}, {Overzier}, {Kauffmann}, {Basu-Zych}, {Leitherer}, {Sembach}, {Martin}, {Rich}, {Schiminovich}, \& {Seibert}}]{heckman11}
{Heckman}, T.~M., {Borthakur}, S., {Overzier}, R., {et~al.} 2011, \apj, 730, 5, \dodoi{10.1088/0004-637X/730/1/5}

\bibitem[{{Hoopes} {et~al.}(2007){Hoopes}, {Heckman}, {Salim}, {Seibert}, {Tremonti}, {Schiminovich}, {Rich}, {Martin}, {Charlot}, {Kauffmann}, {Forster}, {Friedman}, {Morrissey}, {Neff}, {Small}, {Wyder}, {Bianchi}, {Donas}, {Lee}, {Madore}, {Milliard}, {Szalay}, {Welsh}, \& {Yi}}]{hoopes07}
{Hoopes}, C.~G., {Heckman}, T.~M., {Salim}, S., {et~al.} 2007, \apjs, 173, 441, \dodoi{10.1086/516644}

\bibitem[{{Iwata} {et~al.}(2009){Iwata}, {Inoue}, {Matsuda}, {Furusawa}, {Hayashino}, {Kousai}, {Akiyama}, {Yamada}, {Burgarella}, \& {Deharveng}}]{iwata09}
{Iwata}, I., {Inoue}, A.~K., {Matsuda}, Y., {et~al.} 2009, \apj, 692, 1287, \dodoi{10.1088/0004-637X/692/2/1287}

\bibitem[{{Izotov} {et~al.}(2016{\natexlab{a}}){Izotov}, {Guseva}, {Fricke}, \& {Henkel}}]{izotov16b}
{Izotov}, Y.~I., {Guseva}, N.~G., {Fricke}, K.~J., \& {Henkel}, C. 2016{\natexlab{a}}, \mnras, 462, 4427, \dodoi{10.1093/mnras/stw1973}

\bibitem[{{Izotov} {et~al.}(2021){Izotov}, {Guseva}, {Fricke}, {Henkel}, {Schaerer}, \& {Thuan}}]{izotov21}
{Izotov}, Y.~I., {Guseva}, N.~G., {Fricke}, K.~J., {et~al.} 2021, \aap, 646, A138, \dodoi{10.1051/0004-6361/202039772}

\bibitem[{{Izotov} {et~al.}(2016{\natexlab{b}}){Izotov}, {Schaerer}, {Thuan}, {Worseck}, {Guseva}, {Orlitov{\'a}}, \& {Verhamme}}]{izotov16}
{Izotov}, Y.~I., {Schaerer}, D., {Thuan}, T.~X., {et~al.} 2016{\natexlab{b}}, \mnras, 461, 3683, \dodoi{10.1093/mnras/stw1205}

\bibitem[{{Izotov} {et~al.}(2018{\natexlab{a}}){Izotov}, {Schaerer}, {Worseck}, {Guseva}, {Thuan}, {Verhamme}, {Orlitov{\'a}}, \& {Fricke}}]{izotov18}
{Izotov}, Y.~I., {Schaerer}, D., {Worseck}, G., {et~al.} 2018{\natexlab{a}}, \mnras, 474, 4514, \dodoi{10.1093/mnras/stx3115}

\bibitem[{{Izotov} {et~al.}(2018{\natexlab{b}}){Izotov}, {Worseck}, {Schaerer}, {Guseva}, {Thuan}, {Fricke}, \& {Orlitov{\'a}}}]{izotov18b}
{Izotov}, Y.~I., {Worseck}, G., {Schaerer}, D., {et~al.} 2018{\natexlab{b}}, \mnras, 478, 4851, \dodoi{10.1093/mnras/sty1378}

\bibitem[{{Jaskot} {et~al.}(2024{\natexlab{a}}){Jaskot}, {Silveyra}, {Plantinga}, {Flury}, {Hayes}, {Chisholm}, {Heckman}, {Pentericci}, {Schaerer}, {Trebitsch}, {Verhamme}, {Carr}, {Ferguson}, {Ji}, {Giavalisco}, {Henry}, {Marques-Chaves}, {{\"O}stlin}, {Saldana-Lopez}, {Scarlata}, {Worseck}, \& {Xu}}]{jaskot24}
{Jaskot}, A.~E., {Silveyra}, A.~C., {Plantinga}, A., {et~al.} 2024{\natexlab{a}}, \apj, 972, 92, \dodoi{10.3847/1538-4357/ad58b9}

\bibitem[{{Jaskot} {et~al.}(2024{\natexlab{b}}){Jaskot}, {Silveyra}, {Plantinga}, {Flury}, {Hayes}, {Chisholm}, {Heckman}, {Pentericci}, {Schaerer}, {Trebitsch}, {Verhamme}, {Carr}, {Ferguson}, {Ji}, {Giavalisco}, {Henry}, {Marques-Chaves}, {{\"O}stlin}, {Saldana-Lopez}, {Scarlata}, {Worseck}, \& {Xu}}]{jaskot24b}
---. 2024{\natexlab{b}}, \apj, 973, 111, \dodoi{10.3847/1538-4357/ad5557}

\bibitem[{{Kennicutt} \& {Evans}(2012)}]{kennicutt12}
{Kennicutt}, R.~C., \& {Evans}, N.~J. 2012, \araa, 50, 531, \dodoi{10.1146/annurev-astro-081811-125610}

\bibitem[{{Kewley} {et~al.}(2001){Kewley}, {Dopita}, {Sutherland}, {Heisler}, \& {Trevena}}]{kewley01}
{Kewley}, L.~J., {Dopita}, M.~A., {Sutherland}, R.~S., {Heisler}, C.~A., \& {Trevena}, J. 2001, \apj, 556, 121, \dodoi{10.1086/321545}

\bibitem[{{Kroupa}(2001)}]{kroupa01}
{Kroupa}, P. 2001, \mnras, 322, 231, \dodoi{10.1046/j.1365-8711.2001.04022.x}

\bibitem[{{Labb{\'e}} {et~al.}(2023){Labb{\'e}}, {van Dokkum}, {Nelson}, {Bezanson}, {Suess}, {Leja}, {Brammer}, {Whitaker}, {Mathews}, {Stefanon}, \& {Wang}}]{labbe23}
{Labb{\'e}}, I., {van Dokkum}, P., {Nelson}, E., {et~al.} 2023, \nat, 616, 266, \dodoi{10.1038/s41586-023-05786-2}

\bibitem[{{Le Reste} {et~al.}(2024){Le Reste}, {Cannon}, {Hayes}, {Inoue}, {Kepley}, {Melinder}, {Menacho}, {Adamo}, {Bik}, {Ejdetj{\"a}rn}, {J{\'o}zsa}, {{\"O}stlin}, \& {Taft}}]{lereste24}
{Le Reste}, A., {Cannon}, J.~M., {Hayes}, M.~J., {et~al.} 2024, \mnras, 528, 757, \dodoi{10.1093/mnras/stad3910}

\bibitem[{{Leitherer} {et~al.}(2014){Leitherer}, {Ekstr{\"o}m}, {Meynet}, {Schaerer}, {Agienko}, \& {Levesque}}]{leitherer14}
{Leitherer}, C., {Ekstr{\"o}m}, S., {Meynet}, G., {et~al.} 2014, \apjs, 212, 14, \dodoi{10.1088/0067-0049/212/1/14}

\bibitem[{{Leitherer} {et~al.}(1995){Leitherer}, {Ferguson}, {Heckman}, \& {Lowenthal}}]{leitherer95}
{Leitherer}, C., {Ferguson}, H.~C., {Heckman}, T.~M., \& {Lowenthal}, J.~D. 1995, \apjl, 454, L19, \dodoi{10.1086/309760}

\bibitem[{{Leitherer} {et~al.}(2016){Leitherer}, {Hernandez}, {Lee}, \& {Oey}}]{leitherer16}
{Leitherer}, C., {Hernandez}, S., {Lee}, J.~C., \& {Oey}, M.~S. 2016, \apj, 823, 64, \dodoi{10.3847/0004-637X/823/1/64}

\bibitem[{{Leitherer} {et~al.}(2010){Leitherer}, {Ortiz Ot{\'a}lvaro}, {Bresolin}, {Kudritzki}, {Lo Faro}, {Pauldrach}, {Pettini}, \& {Rix}}]{leitherer10}
{Leitherer}, C., {Ortiz Ot{\'a}lvaro}, P.~A., {Bresolin}, F., {et~al.} 2010, \apjs, 189, 309, \dodoi{10.1088/0067-0049/189/2/309}

\bibitem[{{Leitherer} {et~al.}(2011){Leitherer}, {Tremonti}, {Heckman}, \& {Calzetti}}]{leitherer11}
{Leitherer}, C., {Tremonti}, C.~A., {Heckman}, T.~M., \& {Calzetti}, D. 2011, \aj, 141, 37, \dodoi{10.1088/0004-6256/141/2/37}

\bibitem[{{Long} {et~al.}(2024){Long}, {Antwi-Danso}, {Lambrides}, {Lovell}, {de la Vega}, {Valentino}, {Zavala}, {Casey}, {Wilkins}, {Yung}, {Arrabal Haro}, {Bagley}, {Bisigello}, {Chworowsky}, {Cooper}, {Cooper}, {Cooray}, {Croton}, {Dickinson}, {Finkelstein}, {Franco}, {Gould}, {Hirschmann}, {Hutchison}, {Kartaltepe}, {Kocevski}, {Koekemoer}, {Lucas}, {McKinney}, {Nere}, {Papovich}, {P{\'e}rez-Gonz{\'a}lez}, {Pirzkal}, \& {Santini}}]{long24}
{Long}, A.~S., {Antwi-Danso}, J., {Lambrides}, E.~L., {et~al.} 2024, \apj, 970, 68, \dodoi{10.3847/1538-4357/ad4cea}

\bibitem[{{Makan} {et~al.}(2021){Makan}, {Worseck}, {Davies}, {Hennawi}, {Prochaska}, \& {Richter}}]{makan21}
{Makan}, K., {Worseck}, G., {Davies}, F.~B., {et~al.} 2021, \apj, 912, 38, \dodoi{10.3847/1538-4357/abee17}

\bibitem[{{Malkan} {et~al.}(2003){Malkan}, {Webb}, \& {Konopacky}}]{malkan03}
{Malkan}, M., {Webb}, W., \& {Konopacky}, Q. 2003, \apj, 598, 878, \dodoi{10.1086/379117}

\bibitem[{{Martin} {et~al.}(2005){Martin}, {Fanson}, {Schiminovich}, {Morrissey}, {Friedman}, {Barlow}, {Conrow}, {Grange}, {Jelinsky}, {Milliard}, {Siegmund}, {Bianchi}, {Byun}, {Donas}, {Forster}, {Heckman}, {Lee}, {Madore}, {Malina}, {Neff}, {Rich}, {Small}, {Surber}, {Szalay}, {Welsh}, \& {Wyder}}]{martin05}
{Martin}, D.~C., {Fanson}, J., {Schiminovich}, D., {et~al.} 2005, \apjl, 619, L1, \dodoi{10.1086/426387}

\bibitem[{{Mascia} {et~al.}(2024){Mascia}, {Pentericci}, {Calabr{\`o}}, {Santini}, {Napolitano}, {Arrabal Haro}, {Castellano}, {Dickinson}, {Ocvirk}, {Lewis}, {Amor{\'\i}n}, {Bagley}, {Bhatawdekar}, {Cleri}, {Costantin}, {Dekel}, {Finkelstein}, {Fontana}, {Giavalisco}, {Grogin}, {Hathi}, {Hirschmann}, {Holwerda}, {Jung}, {Kartaltepe}, {Koekemoer}, {Lucas}, {Papovich}, {P{\'e}rez-Gonz{\'a}lez}, {Pirzkal}, {Trump}, {Wilkins}, \& {Yung}}]{mascia24}
{Mascia}, S., {Pentericci}, L., {Calabr{\`o}}, A., {et~al.} 2024, \aap, 685, A3, \dodoi{10.1051/0004-6361/202347884}

\bibitem[{{Mason} \& {GLASS}(2018)}]{mason18}
{Mason}, C., \& {GLASS}, B. 2018, in American Astronomical Society Meeting Abstracts, Vol. 231, American Astronomical Society Meeting Abstracts \#231, 226.02

\bibitem[{{Meynet} {et~al.}(1994){Meynet}, {Maeder}, {Schaller}, {Schaerer}, \& {Charbonnel}}]{meynet94}
{Meynet}, G., {Maeder}, A., {Schaller}, G., {Schaerer}, D., \& {Charbonnel}, C. 1994, \aaps, 103, 97

\bibitem[{{Nanayakkara} {et~al.}(2024){Nanayakkara}, {Glazebrook}, {Jacobs}, {Kawinwanichakij}, {Schreiber}, {Brammer}, {Esdaile}, {Kacprzak}, {Labbe}, {Lagos}, {Marchesini}, {Marsan}, {Oesch}, {Papovich}, {Remus}, \& {Tran}}]{nanayakkara24}
{Nanayakkara}, T., {Glazebrook}, K., {Jacobs}, C., {et~al.} 2024, Scientific Reports, 14, 3724, \dodoi{10.1038/s41598-024-52585-4}

\bibitem[{{Overzier} {et~al.}(2010){Overzier}, {Heckman}, {Schiminovich}, {Basu-Zych}, {Gon{\c{c}}alves}, {Martin}, \& {Rich}}]{overzier10}
{Overzier}, R.~A., {Heckman}, T.~M., {Schiminovich}, D., {et~al.} 2010, \apj, 710, 979, \dodoi{10.1088/0004-637X/710/2/979}

\bibitem[{{Overzier} {et~al.}(2008){Overzier}, {Heckman}, {Kauffmann}, {Seibert}, {Rich}, {Basu-Zych}, {Lotz}, {Aloisi}, {Charlot}, {Hoopes}, {Martin}, {Schiminovich}, \& {Madore}}]{overzier08}
{Overzier}, R.~A., {Heckman}, T.~M., {Kauffmann}, G., {et~al.} 2008, \apj, 677, 37, \dodoi{10.1086/529134}

\bibitem[{{Overzier} {et~al.}(2009){Overzier}, {Heckman}, {Tremonti}, {Armus}, {Basu-Zych}, {Gon{\c{c}}alves}, {Rich}, {Martin}, {Ptak}, {Schiminovich}, {Ford}, {Madore}, \& {Seibert}}]{overzier09}
{Overzier}, R.~A., {Heckman}, T.~M., {Tremonti}, C., {et~al.} 2009, \apj, 706, 203, \dodoi{10.1088/0004-637X/706/1/203}

\bibitem[{{Pauldrach} {et~al.}(2001){Pauldrach}, {Hoffmann}, \& {Lennon}}]{pauldrach01}
{Pauldrach}, A.~W.~A., {Hoffmann}, T.~L., \& {Lennon}, M. 2001, \aap, 375, 161, \dodoi{10.1051/0004-6361:20010805}

\bibitem[{{Pellegrini} {et~al.}(2012){Pellegrini}, {Oey}, {Winkler}, {Points}, {Smith}, {Jaskot}, \& {Zastrow}}]{pellegrini12}
{Pellegrini}, E.~W., {Oey}, M.~S., {Winkler}, P.~F., {et~al.} 2012, \apj, 755, 40, \dodoi{10.1088/0004-637X/755/1/40}

\bibitem[{{Pettini} \& {Pagel}(2004)}]{pettini04}
{Pettini}, M., \& {Pagel}, B. E.~J. 2004, \mnras, 348, L59, \dodoi{10.1111/j.1365-2966.2004.07591.x}

\bibitem[{{Planck Collaboration} {et~al.}(2020){Planck Collaboration}, {Aghanim}, {Akrami}, {Arroja}, {Ashdown}, {Aumont}, {Baccigalupi}, {Ballardini}, {Banday}, {Barreiro}, {Bartolo}, {Basak}, {Battye}, {Benabed}, {Bernard}, {Bersanelli}, {Bielewicz}, {Bock}, {Bond}, {Borrill}, {Bouchet}, {Boulanger}, {Bucher}, {Burigana}, {Butler}, {Calabrese}, {Cardoso}, {Carron}, {Casaponsa}, {Challinor}, {Chiang}, {Colombo}, {Combet}, {Contreras}, {Crill}, {Cuttaia}, {de Bernardis}, {de Zotti}, {Delabrouille}, {Delouis}, {D{\'e}sert}, {Di Valentino}, {Dickinson}, {Diego}, {Donzelli}, {Dor{\'e}}, {Douspis}, {Ducout}, {Dupac}, {Efstathiou}, {Elsner}, {En{\ss}lin}, {Eriksen}, {Falgarone}, {Fantaye}, {Fergusson}, {Fernandez-Cobos}, {Finelli}, {Forastieri}, {Frailis}, {Franceschi}, {Frolov}, {Galeotta}, {Galli}, {Ganga}, {G{\'e}nova-Santos}, {Gerbino}, {Ghosh}, {Gonz{\'a}lez-Nuevo}, {G{\'o}rski}, {Gratton}, {Gruppuso}, {Gudmundsson}, {Hamann}, {Handley}, {Hansen}, {Helou}, {Herranz}, {Hildebrandt}, {Hivon}, {Huang}, {Jaffe},
  {Jones}, {Karakci}, {Keih{\"a}nen}, {Keskitalo}, {Kiiveri}, {Kim}, {Kisner}, {Knox}, {Krachmalnicoff}, {Kunz}, {Kurki-Suonio}, {Lagache}, {Lamarre}, {Langer}, {Lasenby}, {Lattanzi}, {Lawrence}, {Le Jeune}, {Leahy}, {Lesgourgues}, {Levrier}, {Lewis}, {Liguori}, {Lilje}, {Lilley}, {Lindholm}, {L{\'o}pez-Caniego}, {Lubin}, {Ma}, {Mac{\'\i}as-P{\'e}rez}, {Maggio}, {Maino}, {Mandolesi}, {Mangilli}, {Marcos-Caballero}, {Maris}, {Martin}, {Martinelli}, {Mart{\'\i}nez-Gonz{\'a}lez}, {Matarrese}, {Mauri}, {McEwen}, {Meerburg}, {Meinhold}, {Melchiorri}, {Mennella}, {Migliaccio}, {Millea}, {Mitra}, {Miville-Desch{\^e}nes}, {Molinari}, {Moneti}, {Montier}, {Morgante}, {Moss}, {Mottet}, {M{\"u}nchmeyer}, {Natoli}, {N{\o}rgaard-Nielsen}, {Oxborrow}, {Pagano}, {Paoletti}, {Partridge}, {Patanchon}, {Pearson}, {Peel}, {Peiris}, {Perrotta}, {Pettorino}, {Piacentini}, {Polastri}, {Polenta}, {Puget}, {Rachen}, {Reinecke}, {Remazeilles}, {Renault}, {Renzi}, {Rocha}, {Rosset}, {Roudier}, {Rubi{\~n}o-Mart{\'\i}n},
  {Ruiz-Granados}, {Salvati}, {Sandri}, {Savelainen}, {Scott}, {Shellard}, {Shiraishi}, {Sirignano}, {Sirri}, {Spencer}, {Sunyaev}, {Suur-Uski}, {Tauber}, {Tavagnacco}, {Tenti}, {Terenzi}, {Toffolatti}, {Tomasi}, {Trombetti}, {Valiviita}, {Van Tent}, {Vibert}, {Vielva}, {Villa}, {Vittorio}, {Wandelt}, {Wehus}, {White}, {White}, {Zacchei}, \& {Zonca}}]{planck20}
{Planck Collaboration}, {Aghanim}, N., {Akrami}, Y., {et~al.} 2020, \aap, 641, A1, \dodoi{10.1051/0004-6361/201833880}

\bibitem[{{Ramambason} {et~al.}(2020){Ramambason}, {Schaerer}, {Stasi{\'n}ska}, {Izotov}, {Guseva}, {V{\'\i}lchez}, {Amor{\'\i}n}, \& {Morisset}}]{ramambason20}
{Ramambason}, L., {Schaerer}, D., {Stasi{\'n}ska}, G., {et~al.} 2020, \aap, 644, A21, \dodoi{10.1051/0004-6361/202038634}

\bibitem[{{Reddy} {et~al.}(2016){Reddy}, {Steidel}, {Pettini}, {Bogosavljevi{\'c}}, \& {Shapley}}]{reddy16}
{Reddy}, N.~A., {Steidel}, C.~C., {Pettini}, M., {Bogosavljevi{\'c}}, M., \& {Shapley}, A.~E. 2016, \apj, 828, 108, \dodoi{10.3847/0004-637X/828/2/108}

\bibitem[{{Robertson}(2022)}]{robertson22}
{Robertson}, B.~E. 2022, \araa, 60, 121, \dodoi{10.1146/annurev-astro-120221-044656}

\bibitem[{{Robertson} {et~al.}(2015){Robertson}, {Ellis}, {Furlanetto}, \& {Dunlop}}]{robertson15}
{Robertson}, B.~E., {Ellis}, R.~S., {Furlanetto}, S.~R., \& {Dunlop}, J.~S. 2015, \apjl, 802, L19, \dodoi{10.1088/2041-8205/802/2/L19}

\bibitem[{{Roy} {et~al.}(2023){Roy}, {Henry}, {Treu}, {Jones}, {Prieto-Lyon}, {Mason}, {Heckman}, {Nanayakkara}, {Pentericci}, {Mascia}, {Brada{\v{c}}}, {Vanzella}, {Scarlata}, {Boyett}, {Trenti}, \& {Wang}}]{roy23}
{Roy}, N., {Henry}, A., {Treu}, T., {et~al.} 2023, \apjl, 952, L14, \dodoi{10.3847/2041-8213/acdbce}

\bibitem[{{Saldana-Lopez} {et~al.}(2022){Saldana-Lopez}, {Schaerer}, {Chisholm}, {Flury}, {Jaskot}, {Worseck}, {Makan}, {Gazagnes}, {Mauerhofer}, {Verhamme}, {Amor{\'\i}n}, {Ferguson}, {Giavalisco}, {Grazian}, {Hayes}, {Heckman}, {Henry}, {Ji}, {Marques-Chaves}, {McCandliss}, {Oey}, {{\"O}stlin}, {Pentericci}, {Thuan}, {Trebitsch}, {Vanzella}, \& {Xu}}]{saldana-lopez22}
{Saldana-Lopez}, A., {Schaerer}, D., {Chisholm}, J., {et~al.} 2022, \aap, 663, A59, \dodoi{10.1051/0004-6361/202141864}

\bibitem[{{Sanders} {et~al.}(2023){Sanders}, {Shapley}, {Topping}, {Reddy}, \& {Brammer}}]{sanders23}
{Sanders}, R.~L., {Shapley}, A.~E., {Topping}, M.~W., {Reddy}, N.~A., \& {Brammer}, G.~B. 2023, arXiv e-prints, arXiv:2301.06696, \dodoi{10.48550/arXiv.2301.06696}

\bibitem[{{Sawant} {et~al.}(2021){Sawant}, {Pellegrini}, {Oey}, {L{\'o}pez-Hern{\'a}ndez}, \& {Micheva}}]{sawant21}
{Sawant}, A.~N., {Pellegrini}, E.~W., {Oey}, M.~S., {L{\'o}pez-Hern{\'a}ndez}, J., \& {Micheva}, G. 2021, \apj, 923, 78, \dodoi{10.3847/1538-4357/ac2c85}

\bibitem[{{Schreiber} {et~al.}(2018){Schreiber}, {Labb{\'e}}, {Glazebrook}, {Bekiaris}, {Papovich}, {Costa}, {Elbaz}, {Kacprzak}, {Nanayakkara}, {Oesch}, {Pannella}, {Spitler}, {Straatman}, {Tran}, \& {Wang}}]{schreiber18}
{Schreiber}, C., {Labb{\'e}}, I., {Glazebrook}, K., {et~al.} 2018, \aap, 611, A22, \dodoi{10.1051/0004-6361/201731917}

\bibitem[{{Stasi{\'n}ska} {et~al.}(2015){Stasi{\'n}ska}, {Izotov}, {Morisset}, \& {Guseva}}]{stasinska15}
{Stasi{\'n}ska}, G., {Izotov}, Y., {Morisset}, C., \& {Guseva}, N. 2015, \aap, 576, A83, \dodoi{10.1051/0004-6361/201425389}

\bibitem[{{Steidel} {et~al.}(2018){Steidel}, {Bogosavljevi{\'c}}, {Shapley}, {Reddy}, {Rudie}, {Pettini}, {Trainor}, \& {Strom}}]{steidel18}
{Steidel}, C.~C., {Bogosavljevi{\'c}}, M., {Shapley}, A.~E., {et~al.} 2018, \apj, 869, 123, \dodoi{10.3847/1538-4357/aaed28}

\bibitem[{{Thomas} {et~al.}(2013){Thomas}, {Steele}, {Maraston}, {Johansson}, {Beifiori}, {Pforr}, {Str{\"o}mb{\"a}ck}, {Tremonti}, {Wake}, {Bizyaev}, {Bolton}, {Brewington}, {Brownstein}, {Comparat}, {Kneib}, {Malanushenko}, {Malanushenko}, {Oravetz}, {Pan}, {Parejko}, {Schneider}, {Shelden}, {Simmons}, {Snedden}, {Tanaka}, {Weaver}, \& {Yan}}]{thomas13}
{Thomas}, D., {Steele}, O., {Maraston}, C., {et~al.} 2013, \mnras, 431, 1383, \dodoi{10.1093/mnras/stt261}

\bibitem[{{Vanzella} {et~al.}(2018){Vanzella}, {Nonino}, {Cupani}, {Castellano}, {Sani}, {Mignoli}, {Calura}, {Meneghetti}, {Gilli}, {Comastri}, {Mercurio}, {Caminha}, {Caputi}, {Rosati}, {Grillo}, {Cristiani}, {Balestra}, {Fontana}, \& {Giavalisco}}]{vanzella18}
{Vanzella}, E., {Nonino}, M., {Cupani}, G., {et~al.} 2018, \mnras, 476, L15, \dodoi{10.1093/mnrasl/sly023}

\bibitem[{{Vanzella} {et~al.}(2019){Vanzella}, {Calura}, {Meneghetti}, {Castellano}, {Caminha}, {Mercurio}, {Cupani}, {Rosati}, {Grillo}, {Gilli}, {Mignoli}, {Fiorentino}, {Arcidiacono}, {Lombini}, \& {Cortecchia}}]{vanzella19}
{Vanzella}, E., {Calura}, F., {Meneghetti}, M., {et~al.} 2019, \mnras, 483, 3618, \dodoi{10.1093/mnras/sty3311}

\bibitem[{{Veilleux} \& {Osterbrock}(1987)}]{veilleux87}
{Veilleux}, S., \& {Osterbrock}, D.~E. 1987, \apjs, 63, 295, \dodoi{10.1086/191166}

\bibitem[{Virtanen {et~al.}(2020)Virtanen, Gommers, Oliphant, Haberland, Reddy, Cournapeau, Burovski, Peterson, Weckesser, Bright, {van der Walt}, Brett, Wilson, Millman, Mayorov, Nelson, Jones, Kern, Larson, Carey, Polat, Feng, Moore, {VanderPlas}, Laxalde, Perktold, Cimrman, Henriksen, Quintero, Harris, Archibald, Ribeiro, Pedregosa, {van Mulbregt}, \& {SciPy 1.0 Contributors}}]{scipy20}
Virtanen, P., Gommers, R., Oliphant, T.~E., {et~al.} 2020, Nature Methods, 17, 261, \dodoi{10.1038/s41592-019-0686-2}

\bibitem[{{Wang} {et~al.}(2019){Wang}, {Heckman}, {Leitherer}, {Alexandroff}, {Borthakur}, \& {Overzier}}]{wang19}
{Wang}, B., {Heckman}, T.~M., {Leitherer}, C., {et~al.} 2019, \apj, 885, 57, \dodoi{10.3847/1538-4357/ab418f}

\bibitem[{{Wang} {et~al.}(2021){Wang}, {Heckman}, {Amor{\'\i}n}, {Borthakur}, {Chisholm}, {Ferguson}, {Flury}, {Giavalisco}, {Grazian}, {Hayes}, {Henry}, {Jaskot}, {Ji}, {Makan}, {McCandliss}, {Oey}, {{\"O}stlin}, {Saldana-Lopez}, {Schaerer}, {Thuan}, {Worseck}, \& {Xu}}]{wang21}
{Wang}, B., {Heckman}, T.~M., {Amor{\'\i}n}, R., {et~al.} 2021, \apj, 916, 3, \dodoi{10.3847/1538-4357/ac0434}

\bibitem[{{Worseck} {et~al.}(2016){Worseck}, {Prochaska}, {Hennawi}, \& {McQuinn}}]{worseck16}
{Worseck}, G., {Prochaska}, J.~X., {Hennawi}, J.~F., \& {McQuinn}, M. 2016, \apj, 825, 144, \dodoi{10.3847/0004-637X/825/2/144}

\bibitem[{{Xu} {et~al.}(2022){Xu}, {Henry}, {Heckman}, {Chisholm}, {Worseck}, {Gronke}, {Jaskot}, {McCandliss}, {Flury}, {Giavalisco}, {Ji}, {Amor{\'\i}n}, {Berg}, {Borthakur}, {Bouche}, {Carr}, {Erb}, {Ferguson}, {Garel}, {Hayes}, {Makan}, {Marques-Chaves}, {Rutkowski}, {{\"O}stlin}, {Rafelski}, {Saldana-Lopez}, {Scarlata}, {Schaerer}, {Trebitsch}, {Tremonti}, {Verhamme}, \& {Wang}}]{xu22}
{Xu}, X., {Henry}, A., {Heckman}, T., {et~al.} 2022, \apj, 933, 202, \dodoi{10.3847/1538-4357/ac7225}

\bibitem[{{Xu} {et~al.}(2023){Xu}, {Henry}, {Heckman}, {Chisholm}, {Marques-Chaves}, {Leclercq}, {Berg}, {Jaskot}, {Schaerer}, {Worseck}, {Amor{\'\i}n}, {Atek}, {Hayes}, {Ji}, {{\"O}stlin}, {Saldana-Lopez}, \& {Thuan}}]{xu23}
---. 2023, \apj, 943, 94, \dodoi{10.3847/1538-4357/aca89a}

\bibitem[{{York} {et~al.}(2000){York}, {Adelman}, {Anderson}, {Anderson}, {Annis}, {Bahcall}, {Bakken}, {Barkhouser}, {Bastian}, {Berman}, {Boroski}, {Bracker}, {Briegel}, {Briggs}, {Brinkmann}, {Brunner}, {Burles}, {Carey}, {Carr}, {Castander}, {Chen}, {Colestock}, {Connolly}, {Crocker}, {Csabai}, {Czarapata}, {Davis}, {Doi}, {Dombeck}, {Eisenstein}, {Ellman}, {Elms}, {Evans}, {Fan}, {Federwitz}, {Fiscelli}, {Friedman}, {Frieman}, {Fukugita}, {Gillespie}, {Gunn}, {Gurbani}, {de Haas}, {Haldeman}, {Harris}, {Hayes}, {Heckman}, {Hennessy}, {Hindsley}, {Holm}, {Holmgren}, {Huang}, {Hull}, {Husby}, {Ichikawa}, {Ichikawa}, {Ivezi{\'c}}, {Kent}, {Kim}, {Kinney}, {Klaene}, {Kleinman}, {Kleinman}, {Knapp}, {Korienek}, {Kron}, {Kunszt}, {Lamb}, {Lee}, {Leger}, {Limmongkol}, {Lindenmeyer}, {Long}, {Loomis}, {Loveday}, {Lucinio}, {Lupton}, {MacKinnon}, {Mannery}, {Mantsch}, {Margon}, {McGehee}, {McKay}, {Meiksin}, {Merelli}, {Monet}, {Munn}, {Narayanan}, {Nash}, {Neilsen}, {Neswold}, {Newberg}, {Nichol}, {Nicinski},
  {Nonino}, {Okada}, {Okamura}, {Ostriker}, {Owen}, {Pauls}, {Peoples}, {Peterson}, {Petravick}, {Pier}, {Pope}, {Pordes}, {Prosapio}, {Rechenmacher}, {Quinn}, {Richards}, {Richmond}, {Rivetta}, {Rockosi}, {Ruthmansdorfer}, {Sandford}, {Schlegel}, {Schneider}, {Sekiguchi}, {Sergey}, {Shimasaku}, {Siegmund}, {Smee}, {Smith}, {Snedden}, {Stone}, {Stoughton}, {Strauss}, {Stubbs}, {SubbaRao}, {Szalay}, {Szapudi}, {Szokoly}, {Thakar}, {Tremonti}, {Tucker}, {Uomoto}, {Vanden Berk}, {Vogeley}, {Waddell}, {Wang}, {Watanabe}, {Weinberg}, {Yanny}, {Yasuda}, \& {SDSS Collaboration}}]{york00}
{York}, D.~G., {Adelman}, J., {Anderson}, John~E., J., {et~al.} 2000, \aj, 120, 1579, \dodoi{10.1086/301513}

\bibitem[{{Yuan} {et~al.}(2021){Yuan}, {Zheng}, {Lin}, {Zhu}, \& {Rahna}}]{yuan21}
{Yuan}, F.-T., {Zheng}, Z.-Y., {Lin}, R., {Zhu}, S., \& {Rahna}, P.~T. 2021, \apjl, 923, L28, \dodoi{10.3847/2041-8213/ac4170}

\bibitem[{{Zhu} {et~al.}(2024){Zhu}, {Yuan}, {Jiang}, {Zheng}, \& {Lin}}]{zhu24}
{Zhu}, S., {Yuan}, F.-T., {Jiang}, C., {Zheng}, Z.-Y., \& {Lin}, R. 2024, \apjl, 974, L20, \dodoi{10.3847/2041-8213/ad7b18}

\end{thebibliography}
\bibliographystyle{aasjournal}
\end{document}